\documentclass[useAMS,usenatbib]{mn2e}

\usepackage{epsfig}
\usepackage{amsmath}
\usepackage{amssymb}
\usepackage{ifthen}
\usepackage{txfonts}
\usepackage{rotating}
\usepackage{url}
\usepackage{varioref}
\usepackage{verbatim}
\usepackage{latexsym}
\usepackage{graphicx}

\DeclareSymbolFont{cmletters}{OML}{cmm}{m}{it}
\DeclareMathSymbol{v}{\mathalpha}{cmletters}{"76}

\usepackage{multirow}



\DeclareSymbolFont{cmletters}{OML}{cmm}{m}{it}

\DeclareMathSymbol{v}{\mathord}{cmletters}{"76}

\def\be{\begin{equation}}
\def\ee{\end{equation}}

\newcommand{\cut}[1]{\hbox{}}

\DeclareSymbolFont{cmletters}{OML}{cmm}{m}{it}
\DeclareMathSymbol{v}{\mathalpha}{cmletters}{"76}
\usepackage{caption}
\usepackage{subcaption}
\usepackage{ifthen}
\usepackage[usenames,dvipsnames]{color}
\usepackage{graphicx}

\newcommand\apj{\rmfamily{ApJ}}%
\newcommand\mnras{\rmfamily{MNRAS}}%
\newcommand\pasj{\rmfamily{PASJ}}%
\defcitealias{bz77}{BZ77}

\usepackage{float}
\usepackage{morefloats}

\usepackage{xcolor}

\usepackage{ulem}

\title[3D GRRMHD Simulations of a Thin Accretion Disk in the MAD state]
{General Relativistic Radiation Magnetohydrodynamic Simulations of Thin Magnetically Arrested Disks}

\author[D.~Morales~Teixeira,
M.~J.~Avara,
J.~C.~McKinney
]
{Danilo Morales Teixeira$^{1,2}$\thanks{\hbox{E-mail: danilo@ita.br~(DMT)}},
Mark J. Avara$^{3,4}$\thanks{Frontiers in Gravitational Wave Astrophysics Postdoctoral Fellow},
Jonathan C. McKinney$^{2,5}$
\\
 $^1$Aeronautics Institute of Technology (IEFM/ITA), Sao Jose dos Campo, Sao Paulo, 12228-900, Brazil \\
 $^2$University of Maryland at College Park, Dept. of Physics, 3114 Physics Science Complex, College Park, MD 20742, USA \\
 $^3$University of Maryland at College Park, Department of Astronomy, 1113 Physical Sciences Complex, College Park, MD 20742, USA\\
 $^4$Center for Computational Relativity and Gravitation, Rochester Institute of Technology, Rochester, NY 14623\\
 $^5$Joint Space-Science Institute,1113 Physical Sciences Complex, College Park, MD 27042, USA
 }

\begin{document}
\date{Received 2013; in original form 2013.}
\pagerange{\pageref{firstpage}--\pageref{lastpage}} \pubyear{2013}
\maketitle

\label{firstpage}

\begin{abstract}

The classical, relativistic thin-disk theory of Novikov and Thorne (NT) predicts a maximum accretion efficiency of 40\% for an optically thick, radiatively efficient accretion disk around a maximally spinning black hole (BH). However, when a strong magnetic field is introduced to numerical simulations of thin disks, large deviations in efficiencies are observed, in part due to mass and energy carried by jets and winds launched by the disk or BH spin. The total efficiency of accretion can be significantly enhanced beyond that predicted by NT but it has remained unclear how the radiative component is affected. In order to study the effect of a dynamically relevant  large-scale magnetic field on radiatively efficient accretion, we have performed numerical 3D general relativistic - radiative - magnetohydroynamic (GRRMHD) simulations of a disk with scale height to radius ratio of $H/R\sim0.1$ around a moderately spinning BH ($a=0.5$) using the code HARMRAD. Our simulations are fully global and allow us to measure the jet, wind, and radiative properties of a magnetically arrested disk (MAD) that is kept thin via self-consistent transport of energy by radiation using the M1 closure scheme.

Our fiducial disk is MAD out to a radius of $\sim16R_g$ and the majority of the total $\sim13\%$ efficiency of the accretion flow is carried by a magnetically driven wind. We find that the radiative efficiency is slightly suppressed compared to NT, contrary to prior MAD GRMHD simulations with an \textit{ad hoc} cooling function, but it is unclear how much of the radiation and thermal energy trapped in the outflows could ultimately escape. 

\end{abstract}


\section{Introduction}
\label{sec_intro}
\newcommand{\MBH}{{M}}
\newcommand{\MBHO}{{M_i}}
\newcommand{\Mdot}{{\dot{M}_0}}
\newcommand{\Mdotedd}{{\dot{M}_{\rm Edd}}}

Black hole (BH) accretion disks are present in Active Galactic Nuclei
(AGNs), X-ray binaries, gamma-ray bursts (GRBs) and tidal disruption
events (TDEs). They are able to convert gravitational potential energy
and BH spin energy into radiation, jets and winds through the stresses
induced via the magnetorotational instability (MRI) \citep{Balbus91,
  Balbus98}, magnetic field threading the disks \citep{BP82} and
magnetic field threading the black hole \citep{BZ77}. \citet{Novikov73}, hereafter NT73, developed a general relativistic model
for thin disks which has been successfully applied to interpret observations of many
astrophysical systems, however, a few important assumptions are inherent to their model. They assumed
all the energy is carried away by the radiation, the
presence of jets and winds were excluded, and radiative
emission approximately ceases inside the inner-most stable circular orbit (ISCO),
where stresses in the NT73 model disappear and a maximally spinning black
hole is then predicted to have an efficiency of 42\%. However, \citet{Gammie99} and \citet{Krolik99} suggested 
that additional
stress inside the ISCO can increase the gravitational
potential energy converted by the disk beyond the NT73 prediction.

General Relativistic Magnetohydrodynamic (GRMHD) simulations of standard and normal (SANE)
disks (untilted and tilted) have found at most $\sim$ 10\% deviations in 
accretion efficiency compared to the predicted value from the
NT73 model \citep{Shafee08, Noble09, Noble10, Penna10, Morales14} and
this difference is likely associated with the amount of magnetic flux
that threads the black hole and disk.

On the other hand, when an accretion disk
accumulates magnetic flux onto the black hole horizon and into the inner disk until the MRI
is quenched and no more large-scale flux can be concentrated inward (magnetic forces pushing
out balance gas forces pushing in), the total accretion efficiency increases far beyond NT73 values. 
This is referred to as a magnetically arrested disk
(MAD) state \citep{Narayan03}. \citet{Avara16} built upon prior
numerical work to study MAD disks in the thin, radiatively efficient
Sub-Eddington regime for the first time. They were able to demonstrate the
self-consistent inward transport of magnetic flux by a thin disk
thereby saturating into the MAD state, and characterized the
radiation, wind, and jet efficiencies, disk structure, and jet power
in such a system. They found that a Sub-Eddington, moderately thin
disk can be as much as twice as radiatively efficient as the NT73 model
would predict. However, \citet{Avara16} made use of an \textit{ ad hoc} cooling
function to maintain a particular disk thickness, and large unknowns in
converting the cooling in those simulations to the dissipation of a
real system called for more realistic numerical experiments which
include self-consistent radiative transport.

GRMHD simulations of thick disks in the MAD
state have found a broad range of high accretion efficiencies, 30\%, for example, for a black hole with
$a/M$=0.5, and 140\% for a black hole with $a/M$=0.99
\citep{tnm11}. \citet{tm12} studied the differences in the jet power
of prograde and retrograde black holes and found accretion
efficiencies of up to $\sim$105\% for the prograde case and $\sim$35\%
for the retrograde case, for a black hole with $|a/M|$=0.9 and \citet{McKinney12a} confirmed these results through simulations of a thicker disk where they found an accretion efficiency of roughly 300\% for the prograde case and roughly 85\% for the retrograde case. These
simulations were the first to demonstrate that the amount of magnetic
flux threading the horizon and inner disk in the MAD state is
determined by the accretion state alone and not the initial magnetic
conditions of the simulations, so long as sufficient large-scale flux is available to be accreted.

In this paper we study the accretion efficiency and behavior of a thin disk in the
sub-Eddington limit, with half-height ($H$) to radius ($R$) ratio of
$H/R\approx0.1$ around a black hole with dimensionless spin 
($a/M$)=0.5, where $a$ is the black hole spin and $M$ the mass of the black hole, and that reaches the MAD state. We perform 
General Relativistic - Radiation - Magnetohydrodynamics (GRRMHD)
simulations and compare our values with the results from the
simulation MADiHR from \citet{Avara16} that used an \textit{ ad hoc} cooling
function to control the scale-height. In our simulations
we made use of the code HARMRAD \citep{McKinney14} which
includes self-consistent radiative transport and solves the GRRMHD equations using the
M1 closure scheme \citep{Levermore86}.

The structure of the paper is as follows: the methodology and
numerical setup are presented in \textsection\ref{sec:goveqns},
results are presented in \textsection\ref{sec:results}, the discussion is presented in \textsection\ref{sec:discuss} and in
\textsection\ref{sec:conclusions} we present our conclusions and look
forward.

\section{Setup of Simulations}
\label{sec:goveqns}

In this section we will describe the initialization of our simulations. Section \ref{sec:thindiskmodel} describes the initial conditions of the
accretion disk, Section \ref{sec:grid} explains choices made for the two grids we have
adopted in our simulations, and in Section \ref{sec:diagnostic} we describe the
equations used for accretion diagnostics.

\subsection{Thin disk model}
\label{sec:thindiskmodel}

We start with a Keplerian accretion disk around
a black hole of mass $M_{BH}$=10$M_\odot$ and spin $a/M$=0.5, and with 
plasma initial conditions satisfying the NT73 solution. The
initial density is tuned using iterations on low-resolution test simulations
in order to reach a quasi steady-state accretion
flow which has $\dot M$=0.4$\dot M_{Edd}$, where $\dot M_{Edd}$ is the
Eddington accretion rate defined as $\dot
M_{Edd}=(1/\eta_{NT})L_{Edd}/c^2$. $\eta_{NT}$ is the NT73
accretion efficiency and $L_{Edd}$ is the Eddington luminosity defined
as

\begin{equation}
L_{Edd}=\frac{4\pi GMc}{\kappa_{es}}\approx 1.3\times 10^{46}\frac{M_{BH}}{10^8M_\odot}erg s^{-1}.
\end{equation}

\noindent For this setup, the NT73 solution predicts $\eta_{NT}\approx$8.6\% and $\dot
M_{Edd}\approx$1.68$\times$10$^{19}$g s$^{-1}$.

The density profile of our disk was given by

\begin{equation}
\rho(r,z)=\rho _0(r)e^{-z^2/(2H^2)},
\end{equation}

\noindent with 

\begin{equation}
\rho_0(r)=\Sigma/(2H),
\end{equation}

\noindent where $\Sigma$ is the surface density, $H$ is the scale height and
their expressions are given by NT73 solution for the inner region
(equation 5.9.10 in NT73) which is radiation pressure dominated.

The initial radial velocity profile of the disk was given by
\begin{equation}
V_r=\alpha _{visc}(H/R)^2r|\Omega|,
\end{equation}
\noindent where $\Omega$ is the angular velocity and $\alpha _{visc}$
was set to 0.5. This viscosity is consistent with empirical measures from prior simulations MAD disks.

As opposed to non-radiative simulations like MADiHR, the lower resolution vertically across the disk
at large radii leads to unstable evolution when radiation hydrodynamics is included. Therefore we
have truncated the accretion disk at $R_{tr1}$=120$R_g$ to avoid
material from these under-resolved regions contaminating the evolution near the black hole. There is an
exponential cutoff for the density so that the radial profile is

\begin{equation}
\rho(r)=\rho(r)e^{(-r/R_{tr1})+1}.
\end{equation}

\noindent Prior MAD simulations show significant structure inside the innermost stable circular orbit (ISCO), but with a `choked' structure, with $H/R$ smaller than just outside the ISCO. We then set initial conditions approximating this with constant $H/R=0.03$ in this region. Figure \ref{fig:fig2a} includes both initial and evolve radial profiles of $H/R$.

The total ideal pressure was given by
$P_{tot}=(\Gamma_{tot}-1)u_{tot}$ with $\Gamma_{tot}$=4/3 and
$u_{tot}$ is the internal energy density with the pressure randomly
perturbed by 10\% to seed the MRI. The accretion disc was surrounded
by an atmosphere with $\rho=10^{-5}(r/R_g)^{-1.1}$, gas internal
energy density $e_{gas}=10^{-6}(r/R_g)^{-5/2}$ and the radiation
energy density and flux were set by the local thermodynamic
equilibrium (LTE) and flux-limited diffusion (McKinney et al. 2014)
with a negligible radiation atmosphere. 

Within the jet and near the axis the rest-mass is low due to the behavior of the accretion inflow/outflow, and both the rest-mass and internal energy are subject to larger fluctuations during the inversion step of the code from conserved quantities to primitive quantities. These fluctuations, both physical and numerical, require floors to ensure numerical stability. We use a numerical ceiling of $b^2/\rho$=300 for RADHR and RADLR, and $b^2/\rho$=100 for RADvHR, as well as floors $b^2/e_{gas}$=10$^9$, and $e_{gas}/\rho$=10$^{10}$.

We have chosen a configuration for the magnetic field of the disk to
be a large scale and poloidal which has a single loop with a transition to a
split monopolar field at $R_{tr2}$. For $r<R_{tr2}$ the $\phi$ component of
the vector potential was given by

\begin{equation}
A_\phi=MAX((r+5)^\nu 10^{40}-0.02,0)(\sin\theta)^{1+h},
\end{equation}

\noindent with $\nu$=1 and $h$=4 while for $r\ge R_{tr2}$ the vector potential was given by

\begin{equation}
A_\phi=MAX((R_{tr2}+5)^\nu 10^{40}-0.02,0)(\sin\theta)^{1+h(R_{tr2}/r)}.
\end{equation}

\noindent The magnetic field was normalized so that there was approximately one MRI wavelength per
half-height $H$ of the disk, resulting in a ratio of gas+radiation pressure to magnetic pressure of $\beta\approx$10, where $\beta\equiv(p_{gas}+p_{rad})/p_{mag}$. In order to investigate
the effects of the field having a transition to a monopolar field we set
$R_{tr2}$=120$R_g$ for simulations RADHR and RADLR, and
$R_{tr2}$=200$R_g$ for RADvHR. 

For the radiation we have assumed that our initial disk has solar
abundance with mass fractions of hydrogen, helium and metals equal to
$X$=0.7, $Y$=0.28 and $Z$=0.02 which gives an electron fraction of
$Y_e=(1+X)/2$ and mean molecular weight of $\bar\mu\approx$0.62. The
expressions for the electron scattering, absorption-mean energy,
bound-free and free-free opacities were the same used by \citet{McKinney15}.

\subsection{Numerical grid}
\label{sec:grid}

In this section we will describe the two grids we set in our
simulations RADHR, RADLR and RADvHR. For RADHR
and RADLR we have used same grid equations from the simulations A0.94BfN40 and A0.94BpN100 for example from \citet{McKinney12a}, while for RADvHR we built a similar grid with
important modifications described below. 

The cell counts for the radial, $\theta$ and $\phi$ directions and the grid type for each of our simulations
are given in table \ref{tbl1}. In all of our simulations we have adopted the same boundary conditions as in \citet{McKinney12a} where the radial grid has $N_r$ cells spanning from $R_{in}\simeq$1.5$R_g\simeq$0.84$R_h$ (where $R_h$ is the horizon radius) to $R_{out}$=10$^5R_g$ with cell size increasing exponentially until $r_{break}$ and then increasing hyper-exponentially. The value of $R_{in}$ is chosen so that there are 7 active grid cells inside the outer horizon, while $R_{in}$ is outside the inner horizon. Outflow boundary conditions are used for the radial boundaries. The $\theta$ grid has $N_\theta$ cells spanning from 0 to $\pi$, with a concentration of cells into the region around and inside the disk. A re-gridding is used rather than higher cell count to better vertically resolve the disk in RADvHR. The disk scale height of $H/R$=0.1 is targeted with the concentration of cells vertically, as in prior work, and transmissive boundary conditions are used in $\theta$. The $\phi$ grid has $N_\phi$ cells spanning uniformly from 0 to 2$\pi$ with periodic boundary conditions. 

\begin{table}
\caption{Grid resolution}
\begin{center}
\begin{tabular}[h]{|l|r|r|r|r|r|r|}
\hline
Simulation & Polar grid & $N_r$ & $N_\theta$ & $N_\phi$ & $T_{stop} (R_g/c)$ \\
\hline
MADiHR & \citet{Avara16}  & 192  & 96 & 208 & 70,000 \\ 
RADLR & \citet{McKinney12a} & 128 & 64 & 32 & 40,000 \\ 
RADHR & \citet{McKinney12a} & 128 & 128 & 64 & 43,000 \\ 
RADvHR & This work & 128 & 128 & 64 & 18,000  \\
\hline
\hline
\end{tabular}
\end{center}
\label{tbl1}
\end{table}

In the results section we will show that in our simulation RADHR the
MAD state has only built up out to $R$=16$R_g$ and for this reason we
have modified the $\theta$ grid in order to better resolve the turbulence which
raises the efficiency of the transport of magnetic flux which should allow more magnetic flux to be transported to the black hole providing conditions to build MAD state beyond 16$R_g$. Our simulation RADvHR has only been run up
to a time 18,000$R_g/c$, as at this resolution the simulations are prohibitively expensive
and algorithmic improvements would be required in order to run
this type of simulation much longer in an affordable manner.

\subsubsection{Polar grid}
\label{secpolar}

\begin{table*}
\caption{Polar grid Parameters}
\begin{center}
\begin{tabular}[h]{|l|r|r|r|r|r|r|r|r|r|r|r|r|r|r|r|r|r|r|r|}
\hline
Simulation & $n_0$ & $c_2$ & $n_2$  & $r_s$  & $r_0$ & $h_3$  & $r_{0j3}$ & $r_{sj3}$ & $n_{j1}$ & $r_{1j}$ & $r_{oj}$ & $r_{sj}$ & $Q_j$ & $n_\theta$ & $h_\theta$ & $r_{sj2}$ & $r_{0j2}$ & $r_{break}$ \\
\hline
RAD(HR,LR) & 1.0 & 1.0 & 6.0 & 40.0 & 40.0 & 0.1 & 40.0 & 0.0 & 0.5 & 30.0 & 30.0 & 40.0 & 1.8 & 5.0 & 0.1 & 8.0 & 3.0  & 200.0 \\ 
RADvHR & 1.0 & 1.0 & 6.0 & 100.0 & 60.0 & 0.1 & 40.0 & 0.0 & 0.7 & 30.0 & 30.0 & 40.0 & 1.7 & 5.0 & 0.1 & 5.0 & 2.0 & 300.0 \\
\hline
\hline
\end{tabular}
\end{center}
\label{tbl2}
\end{table*}

Before we introduce the equations for the new $\theta$ grid, we must
define some variables that are given by

\begin{equation}
T_r(x)=\frac{e^{-1/x}}{e^{-1/x}+e^{-1/(1-x)}},
\end{equation}

\noindent and

\begin{equation}
Trans(X,L,R)=\begin{cases}
0.0 & \text{if } x\ge L,
\\
1.0 & \text{if } x\ge R,
\\
T_r\left(\frac{x-L}{R-L}\right) & \text{if } L<x<R.
\end{cases}
\end{equation}

To concentrate most of the cells in the disk, the expressions for
$S_0$ and $S_2$ that control the cells at small, middle and large
radii, were changed. In our case $S_0$ was changed to

\begin{equation}
S_{0}(r,r_a,r_b)=Trans(\log(r),\log(r_a),\log(r_b)),
\end{equation}

\noindent where the objective of $S_{0}$ is to control the cell size
at middle and large radii. The expression for $S_2$ has been changed
and is given by

\begin{equation}
S_{2}(r,r_a,r_b)=1-S_{0}(r,r_a,r_b),
\end{equation}

\noindent where this expression for $S_2$ controls the cell size at
small and middle radii. Then the new expression for $h_2$ is given by

\begin{eqnarray}
h_2=h_3S_2(r,40,200)+\nonumber \\
S_0(r,40,200)(h^\prime_2(r)S_2(r,200,500)+h^\prime_2(500)S_0(r,200,500)),
\end{eqnarray}

\noindent where

\begin{equation}
h^\prime_2(r)=h_3+\left(\frac{r-r_{sj3}}{r_{0j3}}\right)^{n_{j1}},
\end{equation}

\noindent is the same equation for $h_2$ in \citet{McKinney12a}. The original equation for $\theta_1$ has been
modified and in our case is given by

\begin{equation}
\theta_1=\theta_0S_2(r,20,200)+\theta^\prime_2S_0(r,20,200)
\end{equation}

\noindent where $\theta_0$ is given by

\begin{equation}
\theta_0=\frac{\pi}{2}\left[h_0(2x^{(2)}-1)+(1-h_0)(2x^{(2)}-1)^{n_{\theta 1}}+1\right],
\end{equation}

\noindent with $h_0$ given by

\begin{eqnarray}
h_0=h_3S_2(r,40,200) \nonumber \\
+S_0(r,40,200)[h^\prime_0(r)S_2(r,20,40)+h^\prime_0(500)S_0(r,20,500)]
\end{eqnarray}

\noindent where $h^\prime_0$ is the same expression for $h_0$ in \citet{McKinney12a} and it is given by

\begin{equation}
h^\prime(r)=2-Q_j(r/r_{1j})^{-n_{j2}[(1/2)+(1/\pi)\arctan(r/r_{0j}-r_{sj}/r_{0j})]}.
\end{equation}

\noindent and $\theta^\prime_2$ is given by

\begin{equation}
\theta^\prime_2=\theta^a_2S_0(r,20,40)+\theta^b_2S_0(r,20,40),
\end{equation}

\noindent where $\theta^a_2$ and $\theta^b_2$ are the original
expressions for $\theta_2$ and $T_2$ presented in \citet{McKinney12a} which are given by

\begin{equation}
\theta^a_2=\frac{\pi}{2}\left[h_2(2x^{(2)}-1)+(1-h_2)(2x^{(2)}-1)^{n_{\theta 1}}+1\right],
\end{equation}

\begin{equation}
\theta^b_2=\frac{\pi}{2}\frac{1}{\arctan(h_2/2)}\left[1+\arctan\left(h_2x^{(m_2)}-\frac{1}{2}\right)\right].
\end{equation}

The expression for $\theta_1$ in our case is given by

\begin{equation}
\theta_1=T_0S_2(r,20,200)+\theta^\prime_2S_0(r,20,200),
\end{equation}

\noindent where $T_0$ is the original expression from \citet{McKinney12a}. 

The expression for $\theta_2$ in our case is given by

\begin{equation}
\theta_2=\frac{\pi}{2}\left[h_\theta(2x^{(2)}-1)+(1-h_\theta)(2x^{(2)}-1)^{n_{\theta 2}}+1\right],
\end{equation}

Finally the expression for $\theta$ is

\begin{equation}
\theta=\theta_2S_2(r,1,10)+S_0(r,1,10)\theta_1.
\end{equation}

\noindent The parameters for the polar grid of each of our simulations are listed in table \ref{tbl2}. 

\subsubsection{Convergence}
\label{secconvergence}

As was done in \citet{McKinney12a, McKinney13, McKinney14, McKinney15}, we
tested the convergence quality factors for the MRI in the $\theta$ and
$\phi$ directions ($Q_{\theta,MRI}$, $Q_{\phi,MRI}$) and the turbulence
($Q_{nlm,cor}$). These measure, respectively, the number of grid cells per
MRI wavelength and the number of
cells per correlation wavelength in the radial, $\theta$, and $\phi$
directions, corresponding to the n,l, and m spatial modes. The values were averaged over the time interval
30,000-43,000$R_g/c$ for RADHR, 30,000-40,000$R_g/c$ for RADLR, and 
15,000-18,000$R_g/c$ for RADvHR, and in $\theta$ and $\phi$ at
$r$=10$R_g$. This radius best represents the non-choked part of the accretion
flow that is in the MAD state and inflow quasi-equilibrium for the majority of the simulation run-time. 

The fiducial simulation RADHR has $Q_{\theta,MRI}\sim$110,
$Q_{\phi,MRI}\sim$13, and $Q_{nlm,cor}\sim$11, 13 and 4
respectively. In the magnetically-arrested regions of the disk, however, resolving the MRI
is not as strong a measure of convergence as in simulations of MRI-dominated disks.
 The value of $Q_{nlm,cor}$ in the $\phi$ direction shows
that the turbulence is somewhat under-resolved, especially in this direction, but 
it is encouraging that the turbulent behavior of the MAD disk, which is dominated by 
magnetic Raleigh-Taylor driven modes, is the same as in the better resolved simulations of
\cite{Avara16}. We have also measured the disk scale-height ($H$) per unit MRI
wavelength, $S_d$, where for $S_d<$0.5 the MRI is
suppressed. Initially for our simulation RADHR, $S_d\sim$0.9, and the
time averaged flow has $S_d\sim$0.26 out to $r\sim$16$R_g$, the
simulation RADLR has $Q_{\theta,MRI}\sim$21, $Q_{\phi,MRI}\sim$8 and
$Q_{nlm,cor}\sim$12, 8, 3 and our simulation RADvHR at its end has
$Q_{\theta,MRI}\sim$56, $Q_{\phi,MRI}\sim$13 and $Q_{nlm,cor}\sim$15,
22, 7 and the MAD state built up to 13$R_g$. RADLR provides a useful benchmark for convergence,
 but in light of its lower resolution we do not use it during the analysis presented in the results section below. It is encouraging 
 that all three simulations show similar long-term behavior despite the differences in resolution. 

\begin{figure*}
\centering
\includegraphics[width=7.0in]{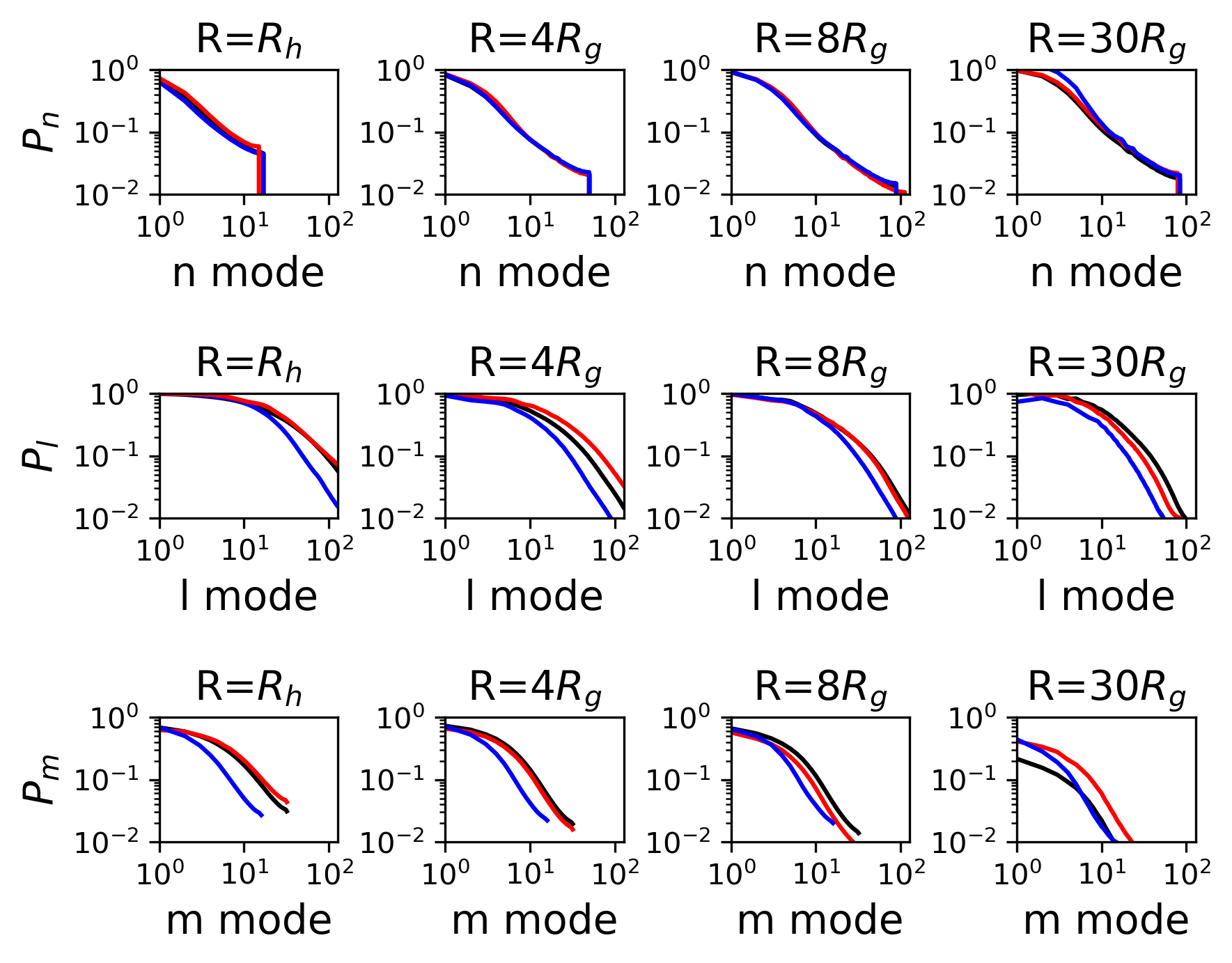}
\caption{Density power spectrum showing the $n$ mode in the top panel, $l$ mode in the middle panel and the $m$ mode in the bottom panel. The black line shows the values for our simulation RADHR, red line for our simulation RADvHR and the blue line for our simulation RADLR.} 
\label{figspec}
\end{figure*}

While there is a significant body of literature devoted to the study of convergence in both global and local accretion disks, the computational expense of simulations with self-consistent radiative transport has prevented the development of similarly trust-worthy convergence criteria. To enable a more detailed comparison to other simulations, we have computed the density power-spectrum for the three modes ($n,l,m$) at four different radii ($R=R_h$, 4$R_g$, 8$R_g$ and 30$R_g$), for all three simulations in this work, and present the density power spectra in Figure \ref{figspec}.

According to the $n$ mode both large and small scales structures in our simulations RADHR and RADvHR have been resolved in the radial direction up to a radius of $\approx$30$R_g$, the $m$ mode demonstrate good convergence in the $\phi$ direction up to a radius of $\approx$10$R_g$ and according to the $l$ mode only large scale structures of the turbulence have been resolved in the $\theta$ direction indicating that the $Q_\theta$ value for simulations with radiative transfer might be different. We are going to carry out a more detailed study to analyze convergence in simulations with radiative transfer in a near future. 

\subsection{Diagnostics}
\label{sec:diagnostic}

In this section we provide the equations used in our analyses to compute the accretion rate, efficiencies, opacities, disk thickness, magnetic flux and magnetic stress. 

The half-angular thickness ($H$) normalized by radius ($R$) is given
by

\begin{equation}
\frac{H}{R}(r,\phi)=\frac{H_0}{R}+\frac{(\int \rho(\theta-\theta_0)^ndA_{\theta\phi})^{1/n}}{(\int \rho dA_{\theta\phi})^{1/n}},
\end{equation}

\noindent where we choose $n$=2 and $H_0/R$=0, unless otherwise
specified, and $dA_{\theta\phi}=\sqrt{-g}dx^{(2)}dx^{(3)}$ is the area
differential. $\theta_0$ is computed as the result of this expression
when one sets $n$=1 and $\theta_0=H_0/R=\pi/2$ on the right hand side.

The mass accretion rate is given by

\begin{equation}
\dot M=\int \rho u^rdA_{\theta\phi},
\end{equation}

\noindent where $\rho$ is the fluid mass density and $u^r$ is the radial
components of the four-velocity 

The accretion efficiency is given by

\begin{equation}
\eta=-\frac{\int(T^r_t+\rho u^r+R_t^r)dA_{\theta\phi}}{[\dot M]},
\end{equation}

\noindent where $T^\mu_\nu$ is the plasma stress-energy tensor,
$R^\mu_\nu$ is the radiation tensor and $[\dot M]$ is the
time-averaged $\dot M$. Different from \citet{Avara16} we normalize by
 values of the accretion rate at every radius, not just $\dot M$ at the 
 horizon, in order to
correct the mass and energy that have been injected during the
evolution to keep the simulation stable.

In these equations $\eta$ is composed of free particle (PAKE, where
the PAKE term is the sum of kinetic plus gravitational terms), thermal
(EN), electromagnetic (EM) and radiation (RAD) components. The jet is
defined as PAKE+EN+EM in regions where the magnetic energy is greater
than the rest mass energy density, $2P_B/\rho>$1 where $P_B$ is the
magnetic pressure. The wind is defined as PAKE+EN and located outside
the jet where $2P_B/\rho<$1 and the flow is outgoing ($u^R>$0). It's
also possible that the wind contains untapped EM and RAD components
that would be converted at unresolved radii distant from the region in
which we reach inflow or outflow equilibrium.

Of central importance to this work is the measurement of magnetic flux
threading the disk and horizon. The magnetic flux on the half
hemisphere of the black hole horizon is given by

\begin{equation}
\Psi_H=0.7\frac{\int^{\theta=\pi}_{\theta=\pi/2}dA_{\theta\phi}B^r}{\sqrt{[\dot M]_H}},
\end{equation}

\noindent where this integral is only carried out over
$\pi/2<\theta<\pi$ to only consider the lower half hemisphere. For
integration over $\theta$=0 to $\theta$=$\pi$/2 the flux is positive
when compared with the equation above for radial magnetic field
strength $B^r$ in Heaviside-Lorentz units.

The magnetic flux in the $\theta$ direction at a radius $r$ and angle
$\theta$ is given by

\begin{equation}
\Psi_\theta(r,\theta)=0.7\frac{\int^r_{r_H}\sqrt{-g}dx^{(1)}dx^{(3)}B^{x^{(2)}}}{\sqrt{[\dot M]_H}},
\end{equation}

\noindent where for simplicity this quantity is calculated using
internal code coordinates $x^\alpha=(t,x^{(1)},x^{(2)},x^{(3)})$. The
total poloidal magnetic flux threading the equatorial plane of the
disk inside some radius is then given by

\begin{equation}
\Psi_{eq}(r)=\Psi_\theta(r,\theta=\pi/2),
\end{equation}

\noindent and thus the total magnetic flux tapping into the rotational
power of the disk and Kerr space-time given by

\begin{equation}
\Psi(r)=\Psi_H+\Psi_{eq}(r),
\end{equation}

\noindent where all definitions of $\Psi$ are integrated over the
entire $\phi$.

Another important characteristic quantity is the absolute magnetic
flux ($\Phi$) which is calculated in the same way as $\Psi$ but by
taking the absolute value of $B^r$ or $B^\theta$,
integrating over all $\theta$, dividing by 2, but not immediately
normalizing by the accretion rate.

\citet{Gammie99} introduced a new normalization for $\Psi$ which has also
been used in this work to normalize $\Phi$ and is given by

\begin{equation}
\Upsilon\approx 0.7\frac{\Phi_r}{\sqrt{[\dot M]_t}},
\end{equation}

\noindent where $\Phi_r(r,\theta)=(1/2)|\int dA_{\theta\phi}|B^r||$
and this already accounts for $\Phi_r$ being in Heaviside-Lorentz
units \citep{Penna10}. Consistent with the notation of \citet{tnm11}, the
magnetic flux in gaussian units is given by

\begin{equation}
\phi_H=\frac{\Phi_H}{\sqrt{\dot Mr^2_gc}},
\end{equation}

\noindent where $\Upsilon=$0.2$\phi_H$.

In the thin disk theory an important measurement is the viscous
parameter $\alpha$, which according to the GR model it is used to
estimate the radial velocity of the flow through the relation

\begin{equation}
v_{r,visc}=-G_1\alpha(H/R)^2|v_\phi|,
\end{equation}

\noindent where $G_1$ is a relativistic correction factor ($G\lesssim$1.5) as defined in \citet{McKinney12a}, which leads to a
measurement of the effective $\alpha$-viscosity given by

\begin{equation}
\alpha_{b,eff}=\frac{v_r}{v_{visc}/\alpha}.
\end{equation}

\noindent The value of $\alpha$ in a disk with a field strength at minimum consistent with 
the saturated MRI is dominated by the magnetic component, 

\begin{equation}
\alpha_b=\alpha_{mag}=-\frac{b^rb_\phi}{p_b+p_{rad}}.
\end{equation}

\noindent The terms due to Reynolds stress are not described in this work
since we have found them, consistent with expectations, to be negligible 
 compared to the Maxwell stress above. We limit our measurement of $\alpha$ to only include disk material
 by selecting plasma where $b^2/\rho<$1, and we volume
average over $\theta$ and $\phi$ with a density weighting. This limits inclusion of the large magnetically-dominated
flux-tubes in the disk when considering characteristics like $\alpha_b$. We can then compare the local measure 
of the stress in dense regions to
$\alpha_{b,eff}$, which captures the effective $\alpha$ taking into account large-scale torques like those driving by 
magnetic winds rather than turbulent stress components.

In order to obtain the optical depth for each instant in time we
compute

\begin{equation}
\tau=\int\rho\kappa_{tot}dl,
\end{equation}

\noindent where for the radial direction $dl=-f_\gamma dr$, $f_\gamma=u^t(1-(v/c)\cos\theta)$, where at large radii $(v/c)\sim 1-1/(u^t)^2$, $\theta$=0 and the integral is from $r_0$=1000 which corresponds to a radius beyond which only transient material would contribute to the optical depth, but a radius the disk wind has reached, to $r$ to obtain $\tau_r(r)$. For the angular direction $dl=f_\gamma rd\theta$, $\theta=\pi$/2 and the integral is from each polar axis towards the equator to obtain $\tau_\theta(\theta)$. The flow's true radiative photosphere is defined when $\tau_r$=1, which corresponds to conservative upper limit to the radius of the photosphere for an observer.

Finally the radiative luminosity is given by

\begin{equation}
L=-\int dA_{\theta\phi}R_t^r,
\end{equation}

\noindent where the luminosity is measured at $r=$50$R_g$ and only for the portion of this shell where
 $\tau_t(r)<$1.

\section{Results}
\label{sec:results}

\begin{figure*}
\centering
\begin{subfigure}{0.45\textwidth}
\centering
\includegraphics[width=\linewidth]{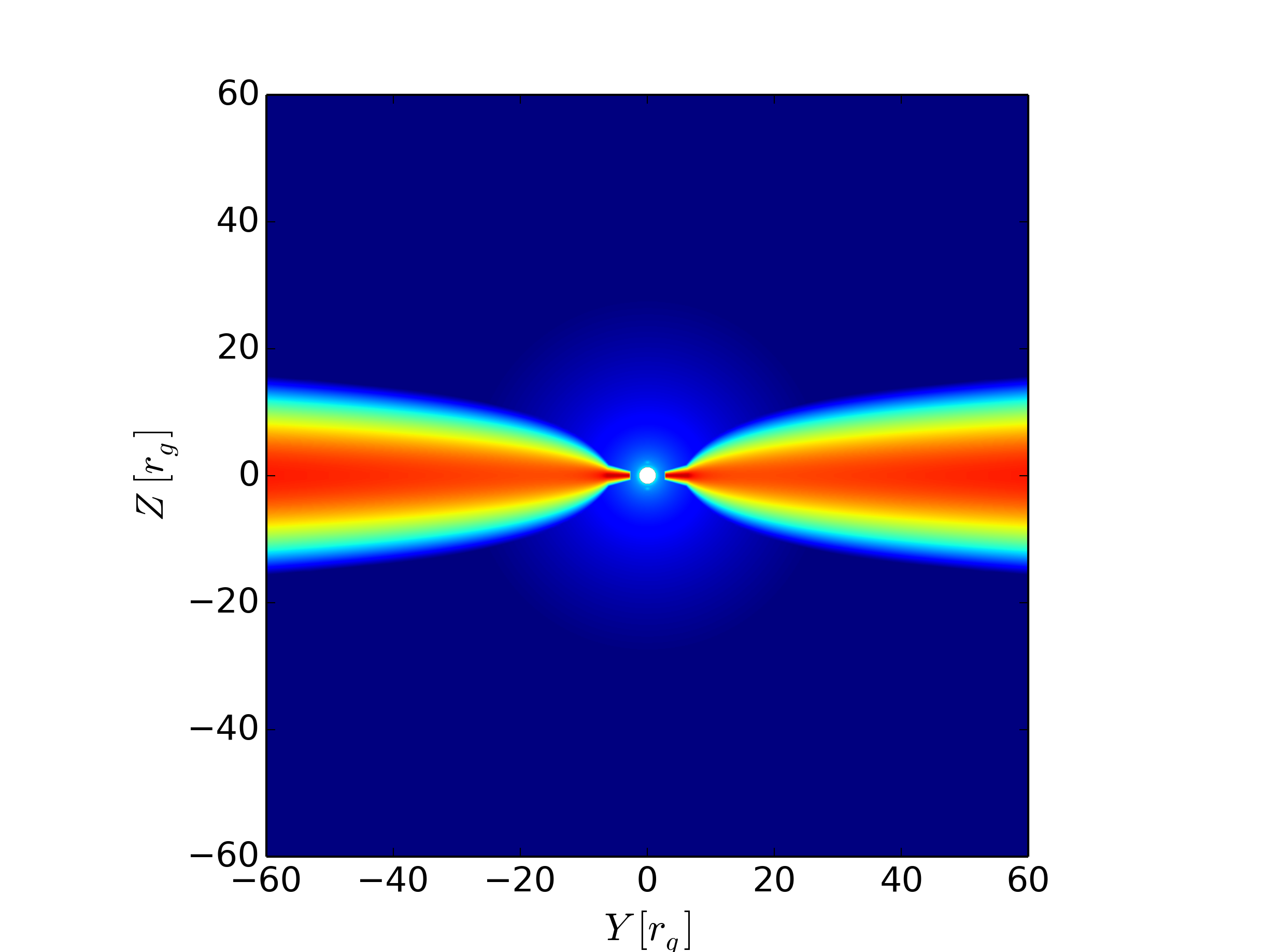} 
\end{subfigure}
\begin{subfigure}{0.45\textwidth}
\centering
\includegraphics[width=\linewidth]{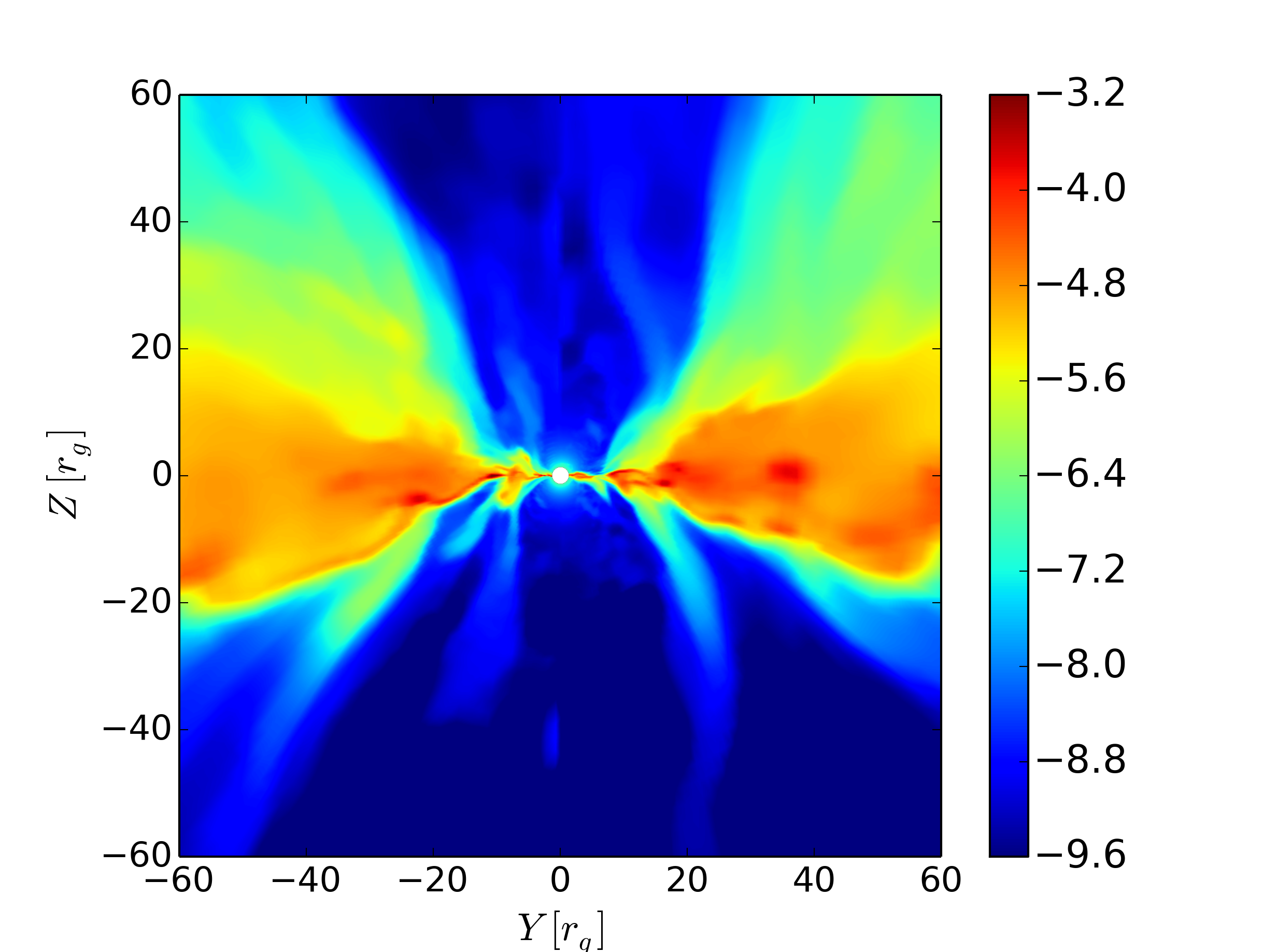}
\end{subfigure}
\caption{Plot of the $X-Z$ slices of the density showing the density at $t$=0$R_g/c$ (left panel) and at $t$=40.000$R_g/c$ (right panel) for our simulation RADHR.}
\label{fig:init_and_evolved_snapshots}
\end{figure*}

In this section we will show how our three simulations evolved from highly magnetized NT73-type
initial conditions into disks that are MAD on the black hole and in the innermost region of the accretion flow. We focus our discussion mainly on the simulation RADHR, our higher-resolution, long-time run, and demonstrate the effect of self-consistent radiative transport on accretion efficiency and general disk behavior as compared to the GRMHD simulations of \citet{Avara16}. We test convergence of the simulations by comparing those run with different grids and resolutions.

Our fiducial simulation RADHR has been run to a final time of $t\approx43,000R_g/c$, by which the inner accretion flow has reached a MAD quasi-equilibrium state where magnetic flux continues to accumulate at the outer edge of the MAD region at $\sim 16R_g$, and the disk is slightly sub-MAD beyond this transition radius where it is MRI dominated. We run the low-resolution analog RADLR to $t=40,000R_g/c$ and we found similar disk structure and evolution.

Figure \ref{fig:init_and_evolved_snapshots} and Figure \ref{fig:RADHRsnapshot} show the three main regimes of behavior in RADHR, culminating in the long-term evolution of the MAD inner disk illustrated in the disk cross-sections. The first stage of evolution involves the rapid evolution of magnetic flux threading the horizon to balance the inner-most accretion flow and dynamical effects of the plunging region inside the innermost stable circular orbit (ISCO), as the system evolves away from analytic NT73 conditions, subject to the dynamically-relevant large-scale poloidal magnetic flux. The second behavioral stage involves mass accretion rate and horizon-integrated magnetic flux growing until $t\sim$17,000$R_g/c$. At that point there is a weak point in the inner disk and a large magnetic disruption occurs which redistributes a lot of poloidal flux into the disk, realizing a state of MAD quasi-equilibrium where the disk is MAD out to the transition radius to MRI-dominated inflow, and magnetic flux continues to slowly accumulate at that radius as it is transported inwards. The final stage, covering roughly the second half of the simulation, is characterized by this quasi-steady state MAD accretion flow, order unity oscillations of the horizon threading flux (consistent with prior numerical findings), and a general accretion structure very similar to the MADfHR simulation in \citet{Avara16}, where only the horizon and inner disk reach the MAD state over the time they were able to run their simulation.

\subsection{Time evolution of the accretion flow and the build up of the MAD state}

In this section we will describe how the MAD state built up in our simulation RADHR by analyzing the time evolution of density,  accretion rate, magnetic flux ($\Upsilon_{\rm H}$), and efficiencies.

\begin{figure*}
\centering
\includegraphics[width=7.0in]{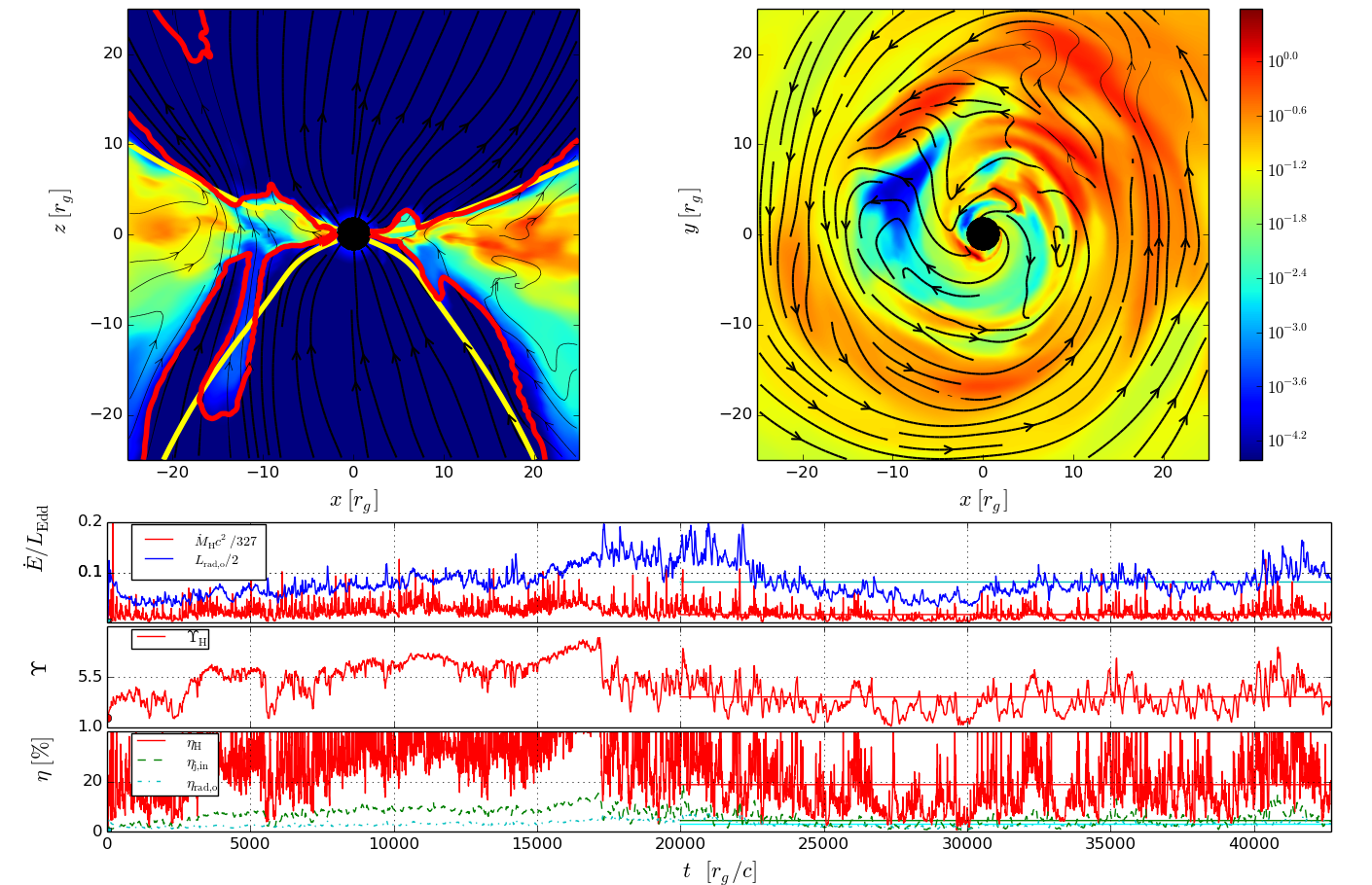}
\caption{ Snapshot of our simulation RADHR at the time $t$=40,000$R_g/c$ where the top two panels represents the $X-Z$ at $y$=0 and $X-Y$ at $z=0$ slices of the density with the red line showing where $b^2/\rho$=1, the yellow line where $\tau$=1 for electron scattering and the black lines with arrows 
	trace field lines where thicker black lines show where field is lightly mass loaded. The top sub-panel shows the mass accretion rate at the 
	horizon (red line) and the luminosity (green line). The middle sub-panel shows the magnetic flux in the horizon normalized so that order unity 
	is a dynamically substantial amount of magnetic flux. The bottom sub-panel shows the accretion efficiencies in the horizon (red line), 
	jet efficiency (green line) measured at 15$R_g$ and radiation (blue line) measure at 50$R_g$.}
\label{fig:RADHRsnapshot}
\end{figure*}

In a movie of frames like Figure \ref{fig:RADHRsnapshot}\footnote{The full movie of our simulation RADHR is 
	available on https://www.youtube.com/watch?v=vbbrCdkcORA} one observes that the flow is highly dynamic with the magnetic Rayleigh-Taylor (RT) instability dominating the flow behavior with periods of high magnetic flux on the black hole (i.e. high $\Upsilon_{\rm H}$), followed by magnetic RT modes resulting in episodes of significant mass accretion and expansion of magnetic flux out into the disk that has been trapped on the horizon by the plunging region of the accretion flow. The large magnetic expulsion events of flux from the horizon into the disk, evolving through the RT instability has analyzed in detail by \citet{Marshall18}, including a detailed study of angular momentum transport by magnetic RT instability in MAD thin disks. 

The disk initial conditions are chosen with enough magnetic flux to ultimately evolve into the MAD state out to large radii, beyond where we could possibly hope to reach inflow equilibrium during the simulation. However, the NT73-type initial conditions necessary to study the disk with self-consistent radiative transport, and attainable resolution constraints, lead to a significant portion of the poloidal magnetic flux in the outer disk being lost through initial transients. The lack of radiation in \citet{Avara16} allowed for initial conditions with more flux at large radii preventing this flux from escaping and so those disks become MAD to a larger radial extent. Despite this loss of flux in the disk, analyzing the suppression factor ($S_d$) of RADHR reveals that at late times the disk still is able to reach the MAD state out to $r=16R_g$. As in MADfHR, this radius grows over time as disk and coronal processes transport more magnetic flux inward, piling up on the outer edge of the MAD region.

The amount of magnetic flux on the horizon increases until steadily until $\sim$17,000$R_g/c$, at which time magnetic pressure, mediated by magnetic reconnection, drives a short episode of flux expulsion into the disk, after which there is a equilibrium time-averaged value of $\Upsilon_{\rm H}$=3.5. The remaining flux on the horizon, saturated at the expected MAD equilibrium level, results in a weak jet. The inclusion of radiative transport does not result in qualitatively different jet structure than non-radiative simulations, lending further support to the finding of \citet{Avara16} that thin accretion disks in the MAD state are able to power a weak, but present, jet. 

We find that the mass accretion rate is also highly variable, like the horizon-threading magnetic flux, for all times, with additional variability associated with evolution away from initial transients up until $\sim$17,000$R_g/c$. After that time the disk is in a quasi-equilibrium state with a mass accretion rate averaged over the time interval 20,000-43,000$R_g/c$, of 0.4$\dot M_{Edd}$. The sharp flickers to high accretion rate are typical of MAD simulations and are associated with accretion of dense clumps of material. The radiative luminosity follows similar variability behavior, but is indicative of a smoothing transfer function of the underlying accretion of material, the source of the radiative energy. The luminosity has a time-averaged value of 0.2$L_{Edd}$. 

Analysis of the time evolution of the efficiencies (total, jet, wind and radiation) has shown that during the entire evolution the wind contributes the largest portion of outgoing energy flux, and overall the system is super-efficient compared to the NT73 predictions. However, in radiative luminosity, the system is under-luminous. The quasi-steady state has a time-averaged total efficiency of 18.6\%, with most energy carried by the particle term. The jet and radiation make similar contributions to the total efficiency with the jet having an efficiency of 4.3\% (mostly in the EM term), and 2.9\% in the radiation. 

Comparing RADHR with the simulation MADiHR (see Figure \ref{fig:efficiencies})  we find that RADHR demonstrates smaller total efficiency, 14\% compared to 20\%, with the largest discrepancy due to the radiative contribution, as one might expect. Self-consistent radiative transport results in only 2.9\% radiative efficiency, well under the MADiHR estimates, and less than half that predicted by NT. 

\subsection{Radial dependence of the accretion flow}

\begin{figure}
\centering
\includegraphics[width=3.2in,clip]{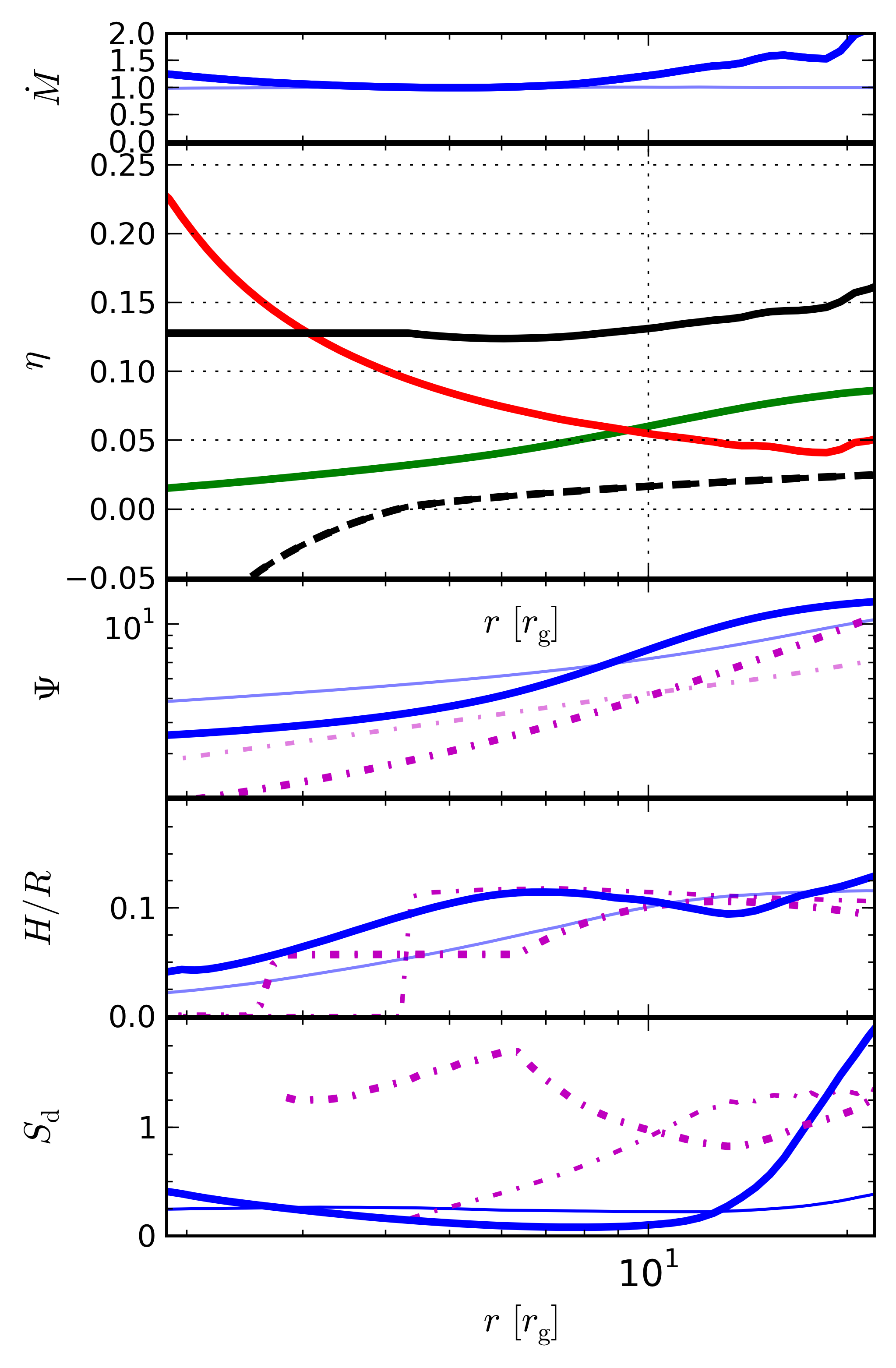}
\caption{Time averaged quantities averaged over the period $t=$[30,000-40,000]$R_g/c$ as function of the radius for our simulation RADHR. The first panel shows the mass accretion rate ($\dot M$), the second panel the energy efficiencies ($\eta$), the following panels the magnetic flux ($\Psi$), the disk scale-height ($H/R$) and the suppression factor ($S_d$). In the plot of $\eta$ the black line corresponds to the total (gas+radiative),  the green line corresponds to the electromagnetic, the red line corresponds to the matter and black dashed line corresponds to radiation components respectively. The magenta line in the plots of $\Psi$, $H/R$ and $S_d$ corresponds to their initial values. The thick lines correspond to our simulation and the thin lines correspond to the simulation MADiHR.}
\label{fig:quantsVsR1}
\end{figure}

Analyzing the mass accretion rate profile (upper panel of Figure \ref{fig:quantsVsR1}) of our simulation RADHR we found that the simulation has reached a quasi-steady state (regions with very similar mass accretion rate) out to a radius of $r\sim15 R_g$, but it is clear from the radial dependence of $\dot M$ that going further out there are larger deviations where the disk is still evolving from initial conditions. The large-scale magnetic flux in these types of simulations enhances these effects due to large-scale coronal flows, large mass transfer from MRI channel modes, and initial radiative evolution since the disk in this simulation does not start in an equilibrium state due to the radiation. Inside the ISCO there is also a deviation in the radial dependence of $\dot M$ due to mass injection associated with the floors which demands careful accounting. It is worth noting that in these and in MADiHR the time-average radial dependence of $\dot M$ near the BH flattens when the number of snapshots over the averaging time period increases. This is largely due to incomplete time-sampling of the high-frequency tail of the variability, characterized by accretion of dense lumps. An additional reason why the radial profile is not flat in simulations which capture the accretion physics near the horizon, is that in order to keep the code stable against numerical instabilities mass and energy must be injected in the domain when the ratio of magnetic energy to rest-mass energy becomes too large.

The additional mass added by the floor in the simulation is removed from the calculation of $\dot M$ where cells have rest-mass values of $b^2/\rho>1$. The profile of MADiHR is much flatter because they time average over a later and longer period, $30,000-70,000R_g/c$ in MADiHR, an \textit{ad hoc} cooling function evolves the disk toward equilibrium faster than the self-consistent processes we capture with full radiative transport, which demands more time to reach thermal equilibrium, and because the higher resolution allows large-scale initial transient modes to break up faster thus having a smaller impact on later evolution (e.g. parasitic modes of the MRI at intermediate and large radii).

The plot of the efficiency (second panel of Figure \ref{fig:quantsVsR1}) and its different components shows an efficiency of roughly 18.6\% at the horizon with most of it in the matter component followed by the electromagnetic part, with the radiation presenting a negative contribution due to the photon capture by the horizon via streaming and also photon advection. Similar efficiencies for all terms but radiation were found in MADiHR where the radiative efficiency was estimated in post-processing from measures of the rate of energy removal by the cooling function prescribed. They accounted for photon capture by light streams intersecting the horizon and photon advection by assuming complete radiative capture inside the photon orbit radius. We now find that this resulted in a total efficiency greater than that measured in our new simulations because their choices lead to an underestimate of the amount of radiation advected by the accretion flow. In addition we now capture the conversion of radiation to thermal and kinetic energy through absorption by wind material. 

For our system with a black hole spinning with $a/M=0.5$, the corresponding value expected for the efficiency according to the NT73 model is 8.2\% and for our simulation RADHR we found a total efficiency of 18.6\% which corresponds to a deviation of 125\%. At the black hole horizon, the amount of the total efficiency in matter, electromagnetic and radiation forms are $\eta^{MAKE}_H$=22.5\%, $\eta^{EM}_H$=1.5\%  and $\eta^{Rad}_H$=-5.4\%.

We measured the jet, wind and radiation efficiencies at $r$=50$R_g$ and we found that at this radius the jet is carrying an efficiency of 4.3\%, the wind is carrying an efficiency of 7.4\% and the radiation an efficiency of 2.9\%, where for the jet 4.1\% is in the electromagnetic form and 0.2\% in the matter while for the wind 4\% is being carried by the electromagnetic part and 3.4\% by the matter. The summed values of the efficiencies carried by the jet, wind and radiation is less than the total efficiency and this difference is because even at $r$=50$R_g$ it's not trivial to guarantee that the wind material will reach infinity and for this reason we have assumed that the wind efficiency is $\eta_{wind}=\eta_{BH}-\eta_{jet}-\eta_{Rad}$ and according to this condition we found that the wind has an efficiency of 11.4\%.

The plot of the amount of magnetic flux (third panel in Figure \ref{fig:quantsVsR1}) in the disk shows that some magnetic flux has been driven outwards since the black hole lost part of its initial flux when compared with the amount of magnetic flux in the beginning of the simulation, compared to MADiHR, where the amount of magnetic flux increased. 

In these simulations with self-consistent radiative transport, we start with a NT73 disk with a scale height of 0.1, but there is no a priori reason to expect that the MAD disk resulting from these initial conditions should persist at this disk width. There was some tuning of initial conditions using test simulations to try to attain a scale height of 0.1 but this result is entirely self-consistent with the mass accretion rate and radiative transport alone. On the other hand, the disk scale height was fixed by a direct cooling function in MADiHR. This is one of the central advantages of including self-consistent radiative transport, which would allow an inherently thermally unstable disk to collapse or puff up if that were the true physical behavior.

The radial profile of the disk shown in Figure \ref{fig:quantsVsR1} and also the density profile of the disk shown in Figure \ref{fig:init_and_evolved_snapshots} show the inner region `choked' by magnetic compression, where the disk scale height is reduced to $\sim0.04$. In the inner edge of the disk the scale height grows to 0.1 at $r$=5$R_g$ and then to 0.12 at $r$=20$R_g$. Beyond that radius we don't show the values because the disk hasn't reached thermal equilibrium in these locations. For comparison, the simulation MADiHR had $H/R \sim0.05$ near the black hole and 0.1 throughout the rest of the disk. Thus there is significant similarity in disk structure as a function of radius in simulations with radiative transfer and those using only an \textit{ad hoc} cooling function. This gives additional credibility to the findings taken from the much longer simulations with \textit{ad hoc} cooling since artificial cooling and the associated choices for cooling timescale are arguably least trustworthy in the plunging region and very inner disk.

The solution of \citet{Shakura76} for the inner region of the disk is radiation pressure dominated and thermally unstable. \citet{Piran78} found a stability criteria for thin disks based on the dependence of the cooling and dissipative terms with the surface density since they are expected to have different slopes and this has been confirmed by \citet{Mishra16}. Unlike these predictions, and similar to the findings of \citet{Sadowski16c,Sadowski16d}, our simulations exhibit thermally stable disks, likely due to the large amount of magnetic flux which can provide a magnetic pressure support strong enough to keep the disk stable. Over the $30,000-40,000R_g/c$ time period for RADHR we found that the disk has reached thermal equilibrium only in the inner parts, not having enough time to reach an equilibrium state further out. We determine that a given radius has reached thermal equilibrium when that part of the disk has reached inflow/outflow quasi-equilibrium and the scale height, in part governed by vertical radiative transport and pressure support evolving under temperature evolution, stabilizes to roughly a fixed value.

Finally the plot of the suppression factor ($S_d$) shows that initially our disk is not MAD (MAD region has $S_d<$0.5) and at the end our simulation has reached the MAD poloidal flux limit out to a radius of 16$R_g$, while the simulation MADiHR was already in the MAD state up to a radius of 30$R_g$.

\subsection{Time evolution of the disk thickness and magnetic flux}

In this section we will describe the time evolution of the accretion flow following the results analyzing how the disk scale-height, magnetic flux and effective viscosity evolve.

\begin{figure}
\centering
\includegraphics[width=3.2in,clip]{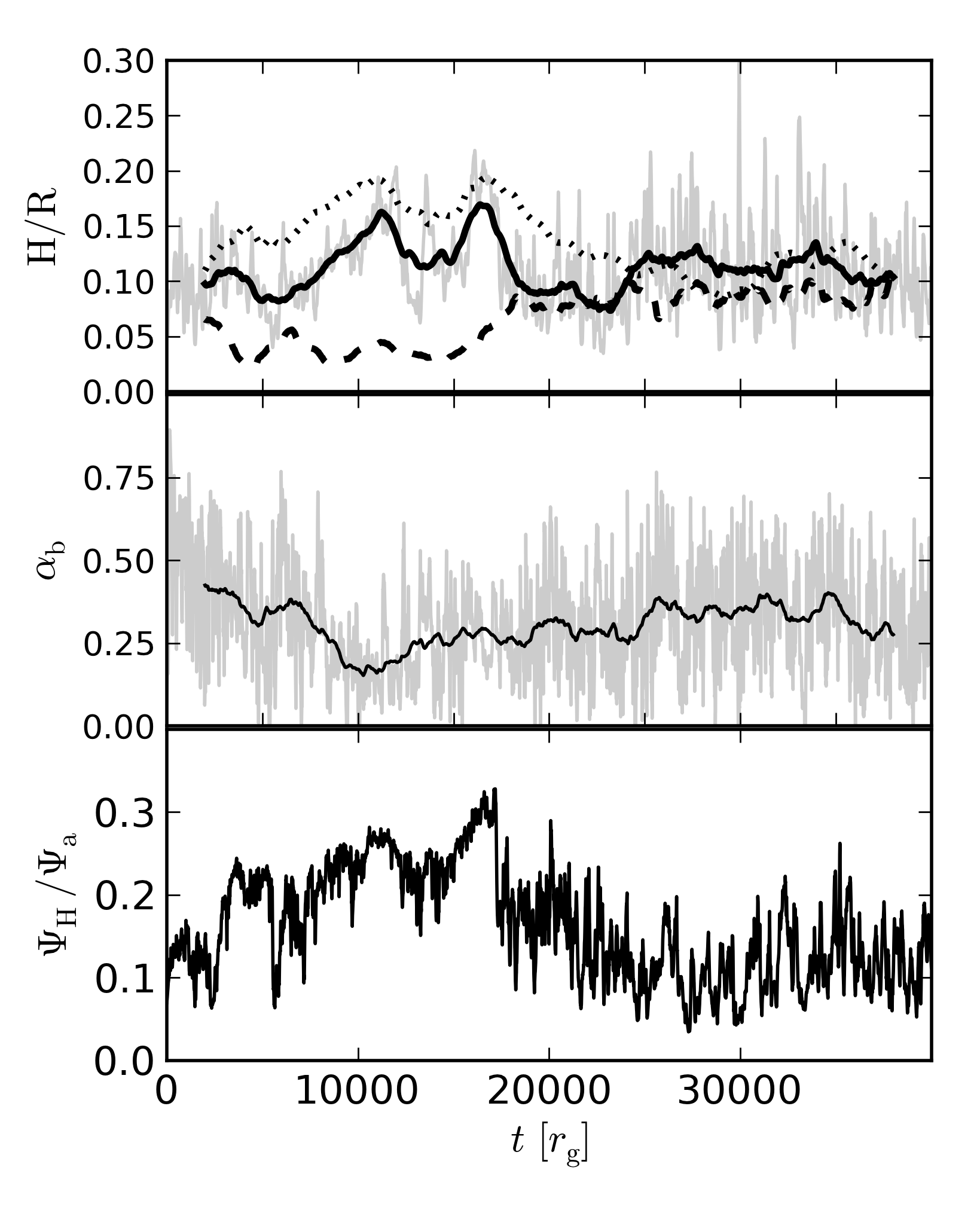}
\caption{Top: Disk scale-height as function of time where the solid line corresponds to the values for 10$R_g$ (lines for all radii are smoothed, with the 10$R_g$ line duplicated without smoothing in gray), the dotted and dashed lines correspond to the values of $H/R$ at the ISCO and 20$R_g$. Middle: Magnetic $\alpha$ measured at 10$R_g$ given by the solid black line (full, un-smoothed data in gray). Bottom: Amount of magnetic flux on the horizon normalized by the amount of magnetic flux contained in the portion of the disk which reaches one inflow time by the end of the simulation, i.e. the total magnetic flux available for the inner disk and horizon.}
\label{fig:quantsVsT}
\end{figure}

\begin{figure}
\centering
\includegraphics[width=3.4in,clip]{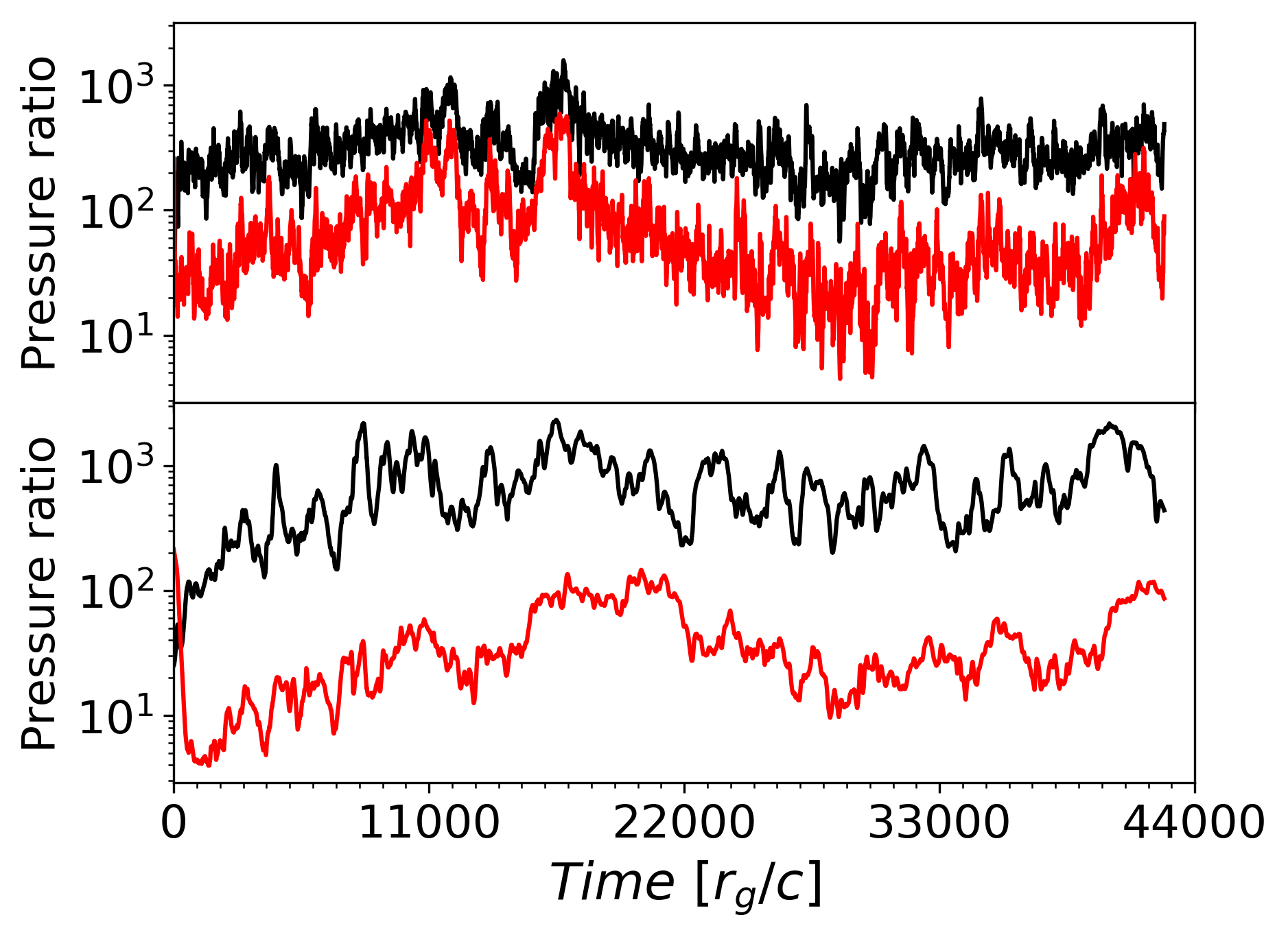}
\caption{Pressure ratios in RADHR, density weighted over the disk and corona regions, and averaged in $\phi$ and time. The black lines refers to the magnetic pressure over gas pressure ratio and red lines to the radiation pressure over gas pressure ratio respectively. The top panel refer to the averaged values over the radial range between the ISCO and 15$r_g$ and the bottom panel refers to the the same measures radially averaged between 20-30$r_g$.}
\label{figpress}
\end{figure}

The time evolution of the disk scale-height (upper panel of Figure \ref{fig:quantsVsT} shows that the disk that started from a non-radiative equilibrium configuration has reached pressure equilibrium up to a radius of 20$R_g$, that the values of $H/R$ at three different locations have converged to roughly the same value of $H/R\approx$0.1, and also demonstrates that during its time evolution the disk remained stable against the collapse due to the radiative loses since the scale height of the disk in the inner region is greater than its initial value. We have also confirmed that during the initial evolution of the system the disk became magnetically pressure dominated demonstrating that the magnetic pressure is supporting the disk against gravitational collapse, which is similar to the finding of \citet{Sadowski16c,Sadowski16d}. The time evolution of the magnetic and radiation to gas pressure ratios is shown in Figure \ref{figpress}

Measuring the absolute flux threading the horizon as a function of time, included in Figure \ref{fig:RADHRsnapshot} and Figure \ref{fig:quantsVsT} for simulation RADHR, we found that the black hole gains a significant quantity of poloidal magnetic flux across the first half of the simulation, in the initial
transitory phase due to large scale MRI channel modes that survive until the $\phi$-symmetry of the field is eaten away by parasitic and other modes. Then, similar to MADiHR, during the early evolution we observe the two modes of flux accumulation as it is transported both through the disk and along the disk surfaces through a coronal-type mechanism. Similar to the build-up of flux from sub-MAD to MAD conditions on the horizon seen in MADfHR of \citet{Avara16} there is a steady build-up of flux on the horizon in RADHR until at $\sim17,000R_g/c$ where there is more flux on the horizon than the disk can confine, and a significant portion of the flux on the horizon finds a low pressure weak spot in the previously rather axisymmetric disk and reconnects, pushing flux out into the disk through that point. 

When a section of the disk reaches the MAD state in such a way the flux the flux does not diffuse away it is expected that the value of $\alpha_{\rm b}$ to saturate to its maximum value according to Equation 34. This is confirmed in our simulation where according to the plot of the time evolution of $\alpha_{\rm b}$ in Figure \ref{fig:quantsVsT} where it is possible to observer that by the time $t$=20,000$R_g/c$ the value of effective $\alpha_{\rm b}$ is roughly 0.25. 
 
\subsection{Luminosities}

\begin{figure}
\centering
\includegraphics[width=3.2in,clip]{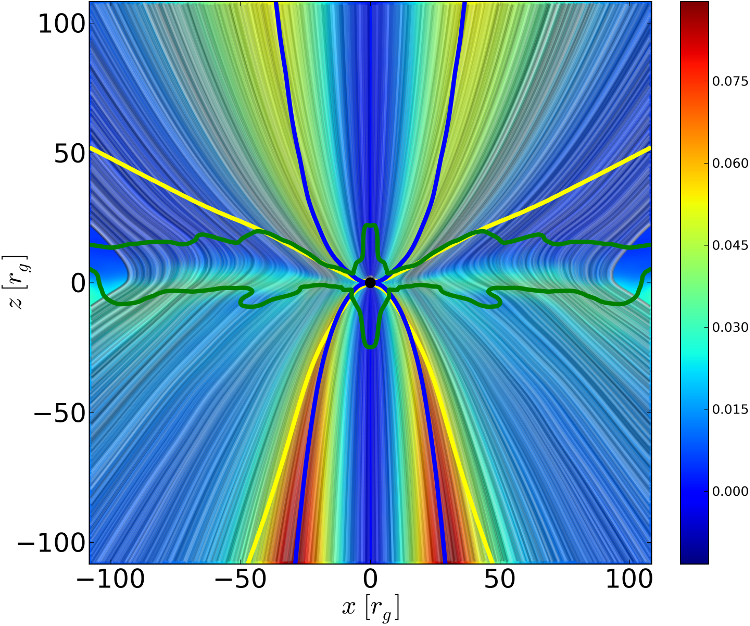} 
\includegraphics[width=3.2in,clip]{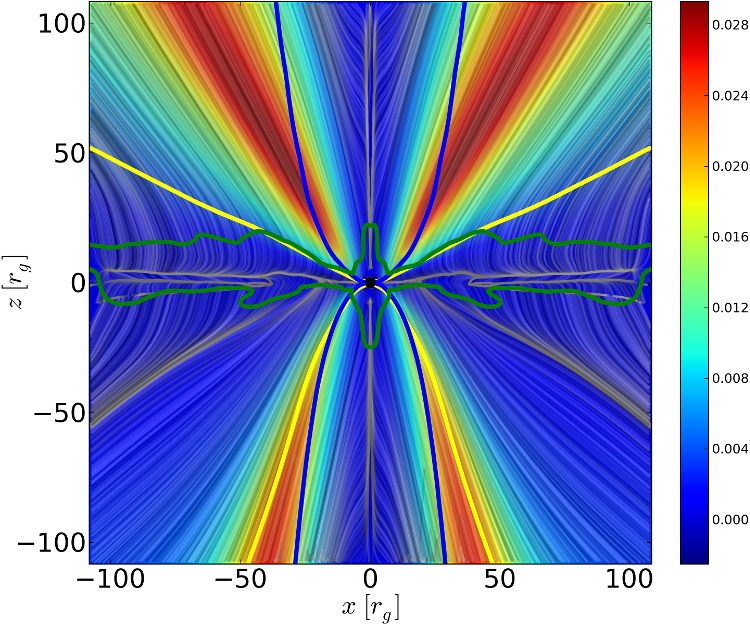}
\caption{Time-$\phi$ averaged electromagnetic luminosity per unity angle ($\partial_\theta L_{EM})/(\dot M_Hc^2)$ (top panel) and radiation flux and radiative luminosity (bottom panel) with the blue, yellow and green lines showing $\phi$-averaged values where $b^2/\rho$=1, $\tau_{es}$=1 and $u^r$=0 respectively.}
\label{fig:fig2a}
\end{figure}

In Figure \ref{fig:fig2a} we show the time-$\phi$ averaged structure of the radiative and EM fluxes for our simulation RADHR, where the average is over the time period 20,000$R_g/c$ to 40,000$R_g/c$, to demonstrate the angular distribution of this outgoing energy, the primary origins in the accretion inflow/outflow, and the distribution of reprocessing. 

The profile of the EM flux shows that most of the outflowing EM energy is directed towards the south and is less isotropically distributed in $\theta$, while the wind pointing to the north direction shows a more uniform distribution. The electron scattering surface (yellow line) has a narrower opening angle in the south direction, due to the substantial quantity of material blocking the radiation at intermediate ($\sim$10-100$R_g$).

The radiation flux along jets in both directions also exhibits a non-uniform distribution having a more enhanced emission in the north side in a opening angle roughly similar to the EM emission. The disk seems to spontaneously break the north-south symmetry of the emission and then the larger radiation flux to the north may maintain that asymmetry for long periods of time. It is not clear if the asymmetry would persist if the RADHR simulation was run longer, but we see this present to some degree in each of our RAD-GRMHD simulations and a weaker asymmetry was seen in some simulations of \citet{Avara16} as well suggesting the magnetic wind may be playing a role in the observed behavior. This kind of asymmetry has also been reported by \citet{Narayan16} when computing the radiative emission from their simulation of a supercritical accretion flow. In their case the asymmetry was also due to the fact M1 scheme needs improvements to properly handle the radiation field in the funnel region. 

A zoom-in in the luminosities shows that most of the emission comes mainly from the region inside $r$=15$R_g$ which corresponds to the region that reached equilibrium. These two images also show a large deviation from the profile expected according to the NT73 model. The NT73 model does not include the presence of jets and winds and for this reason a more uniform distribution would be expected. Our results show that the presence of jets and winds has important effects since they provide a very important contribution to the total luminosity. 

\subsection{Resolution effects} 

In this section we will demonstrate how the resolution affects time evolution of the simulations RADLR and RADvHR, compared with to the fiducial simulation RADHR. We will start describing the time evolution of RADLR which has been run up to a time of 40,000$R_g/c$. Then, we will describe RADvHR, which was run until 18,000$R_g/c$, as long as resources allowed.

 \begin{figure*}
\centering
\includegraphics[width=7.0in]{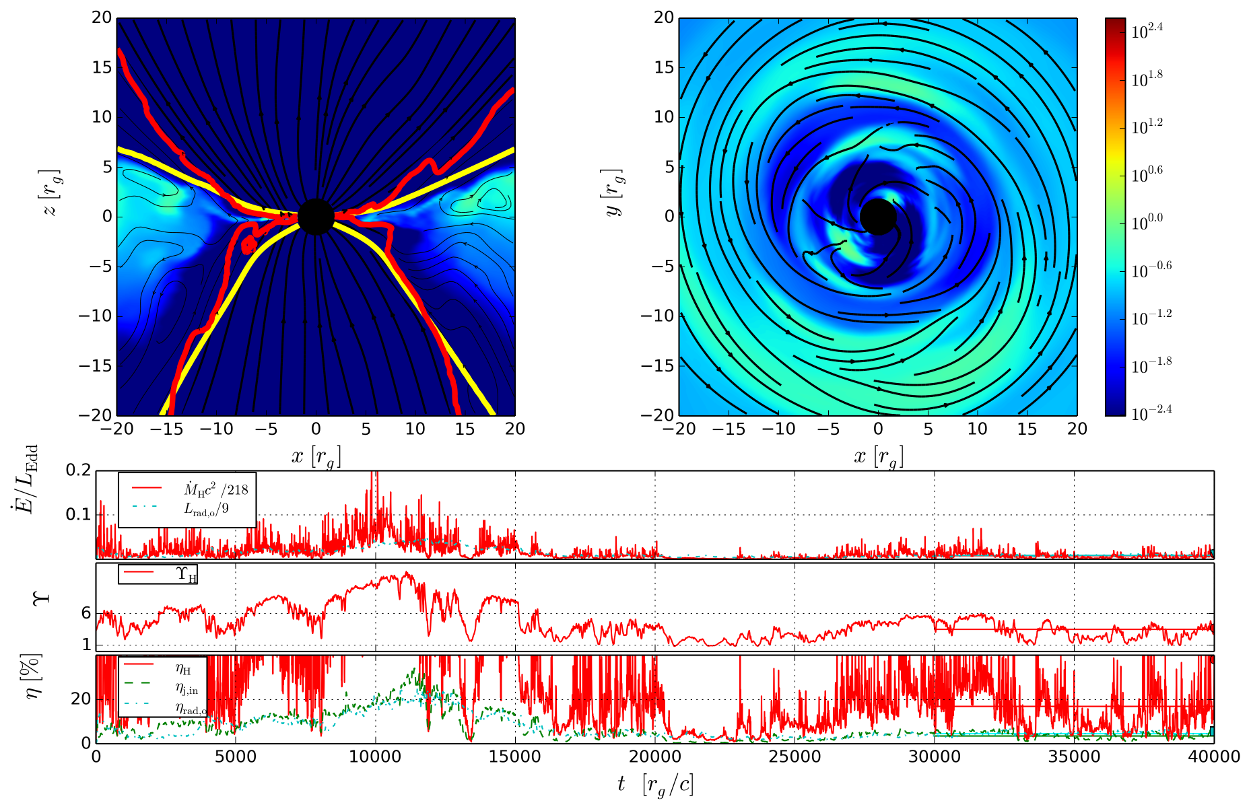}
\caption{Snapshot of our simulation RADLR at $t=$40,000$R_g/c$ similar to Fig~2.} 
\label{fig:RADLR-snapshot}
\end{figure*}

The time evolution of the mass accretion rate of our simulation RADLR (first plot of Figure \ref{fig:RADLR-snapshot}) has increased in the first quarter of the simulation up to a time 12,000$R_g/c$ with the accretion rate decreasing after this time followed by a minimum during the time interval 20,000-27,000$R_g/c$. During the last quarter of evolution, RADLR has an accretion rate exhibiting weak variability, where it is possible to observe magnetic RT modes. It has a final time-averaged mass accretion rate of 0.2$\dot M_{Edd}$, and the radiation luminosity is 0.1$L_{Edd}$. Due to the poor resolution of this simulation, the quasi-steady state seems only to be observed in the last quarter of evolution in time. These three simulations, especially when comparing to MADiHR which had higher $\phi$-resolution, demonstrates the necessity of angular resolution in quickly reaching steady state for accretion simulations starting with strong magnetic fields and axisymmetry.  

The time evolution of the magnetic flux (second plot in Figure \ref{fig:RADLR-snapshot}) on the black hole ($\Psi_H$), similar to RADHR, shows episodes of high magnetic flux indicating that the flow is highly dynamic with magnetic RT instability dominating the accretion flow behavior. Large RT modes allow episodes of significant accretion and release of horizon-trapped magnetic flux. At the end of the simulation RADLR, the black hole has $\Psi\approx$3.5 which is very similar to the value of RADHR. However, flux levels differ in the disk at $r$=10$R_g$ where RADLR has $\Psi$=1.2, versus $\Psi$=2.1 for RADHR. This different distribution of flux leads to the MAD state in RADLR only existing out to a radius of 13$R_g$.

The total efficiency of our simulation RADLR has a deviation of 106\% when compared with the expected value of the NT73 model where the radiative 4.6\% against the 2.9\% for RADHR demonstrating that low resolution simulations the radiative is over estimated indicating that more radiation is absorbed. The jet of RADLR has an efficiency of 3.1\% compared to the 4.3\% for RADHR demonstrating that less rotational energy has been extracted from the black hole as expected since there is less amount of magnetic flux in this region.

 \begin{figure*}
\centering
\includegraphics[width=7.0in]{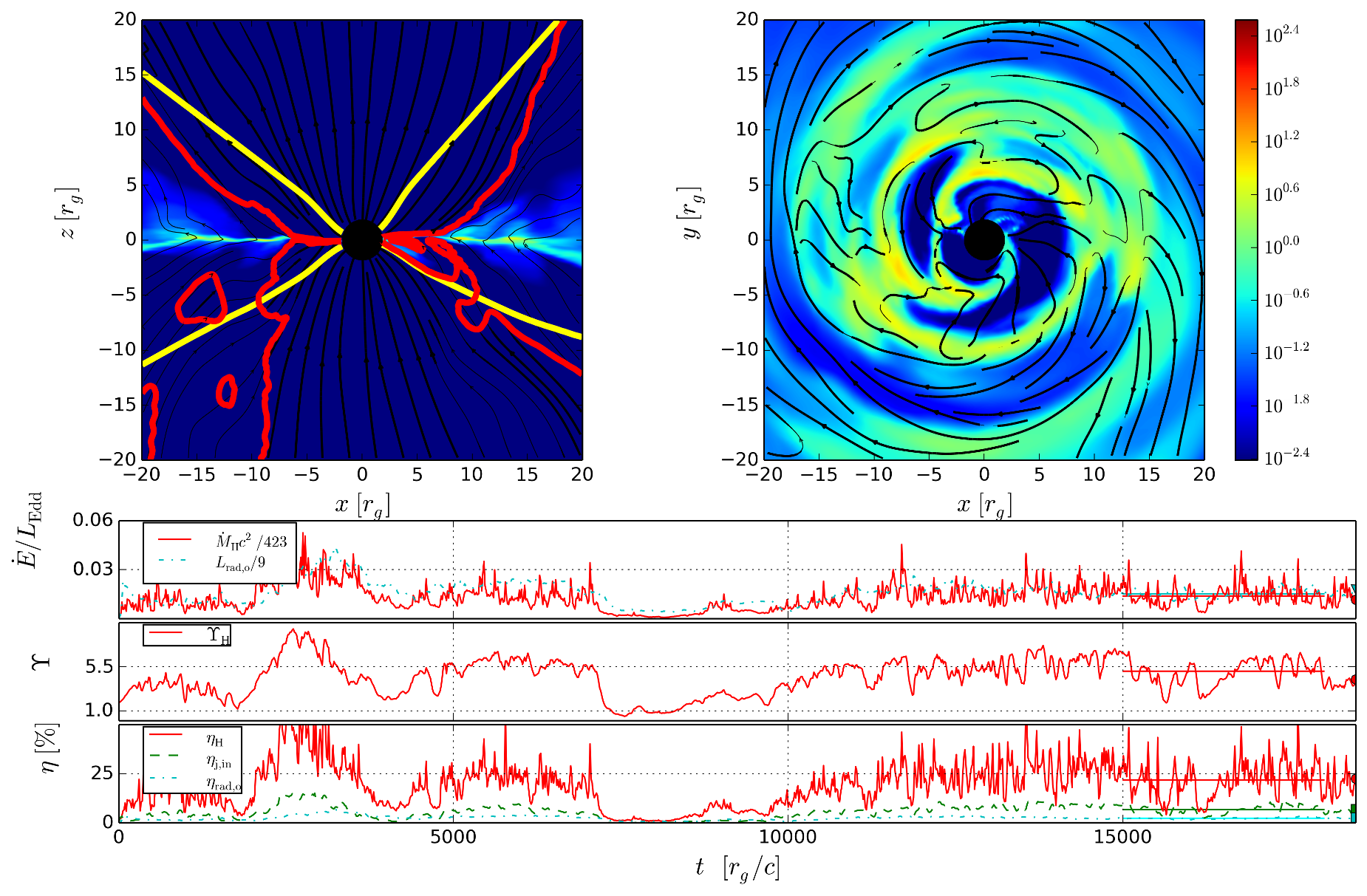}
\caption{Snapshot of our simulation RADvHR at $t=$18,000$R_g/c$ similar to Fig~2.} 
\label{fig:RADvHR}
\end{figure*}

The time evolution of the mass accretion rate for our simulation RADvHR (first plot of Figure \ref{fig:RADvHR}) seems to indicate that the inner accretion flow reached a quasi-steady state after a time of 11,000$R_g/c$ with an averaged mass accretion rate of 0.5$\dot M_{Edd}$ which is very close to the target value and an averaged radiation luminosity of 0.1$L_{Edd}$. Compared to the time evolution of RADHR, we found that RADvHR has reached the quasi-steady state faster.

The time evolution of the magnetic flux (second plot of Figure \ref{fig:RADvHR}) shows that during the time period 7,000-10,000$R_g/c$ the black hole lost a significant amount of magnetic flux reaching a quasi-steady state behavior after this period with an averaged magnetic flux value of $\Psi\approx$4.9.

Due to the better resolution in the polar grid in our simulation RADvHR evolves away from initial transients sooner, as expected. We find a total efficiency of 21.3\%, which corresponds to a deviation of 160\% compared to the expected value from the NT73 model, where in this case the jet has an efficiency of 5.6\%, which compared to RADHR is more powerful and the radiation has an efficiency of 2.4\% compared to the2.9\% in RADHR demonstrating that less material had absorbed the radiation in under-resolved regions that could be located close to the disk.

We have found that resolution has affected the time evolution of the $\tau_{es}$=1 line for electron scattering as well. In RADLR we measure a smaller opening angle in the south direction, possibly due to more material absorbing radiation in under-resolved regions with RADvHR trending in the opposite direction since this simulation has the larger opening angle in the north side. We should point out that since RADvHR hasn't been evolved long enough we cannot guarantee this behavior would be observed at very late times. 

 \begin{figure}
\centering
\includegraphics[width=3.2in,clip]{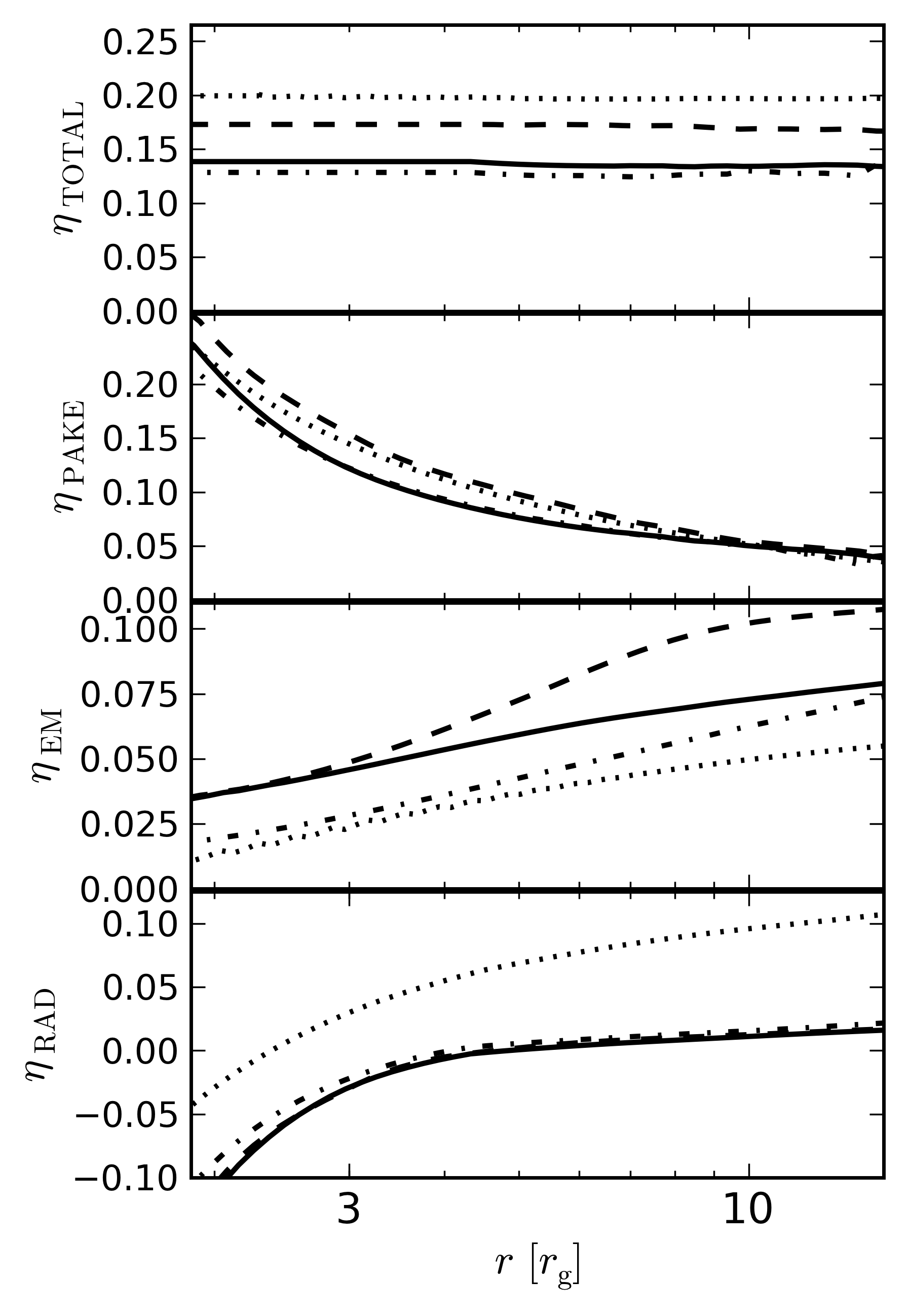}
\caption{Plot of the efficiencies where the plots show the total efficiency, the efficiency carried in the matter form, efficiency carried in the electromagnetic form and the efficiency carried by the radiation respectively. The continuous line corresponds to the simulation RADHR, the dashed line is for the simulation RADvHR, the dashed-dotted line is for the simulation RADLR and the dotted line is for the simulation MADiHR. This plot only shows values inside the radius where the accretion flow of our simulation RADHR has reached thermal equilibrium. The data from simulations RADHR, RADLR and MADiHR were averaged over the time period 30,000-40,000$R_g/c$ while the simulation RADvHR was averaged over the time period 15,000-18,000$R_g/c$.} 
\label{fig:efficiencies}
\end{figure}

The total efficiency and its different components (PAKE, EM and RAD) versus radius averaged over the time period 12,000-18,000$R_g/c$ show that the simulation MADiHR has larger total and radiative efficiencies compared to these RAD simulations. However, the non-radiative components in MADiHR are similar to radiative simulations, suggesting the large difference in total efficiency is the lack of self-consistent radiative transfer. RADHR, RADLR, and RADvHR have similar efficiencies to one another, with the latter demonstrating the highest total efficiency. Figure \ref{fig:efficiencies} presents the values averaged over the time period 30,000-40,000$R_g/c$ for RADHR, RADLR and MADiHR, and during the time period 15,000-18,000$R_g/c$ for RADvHR where one can see that even at the late times this behavior doesn't change, with MADiHR still demonstrating the largest total efficiency. The EM component of RADvHR has is largest during this time period, compared to that component of RADvHR and RADHR, which show similar values inside the plunging region, and this behavior doesn't change at late times.

\begin{figure}
\centering
\includegraphics[width=3.2in,clip]{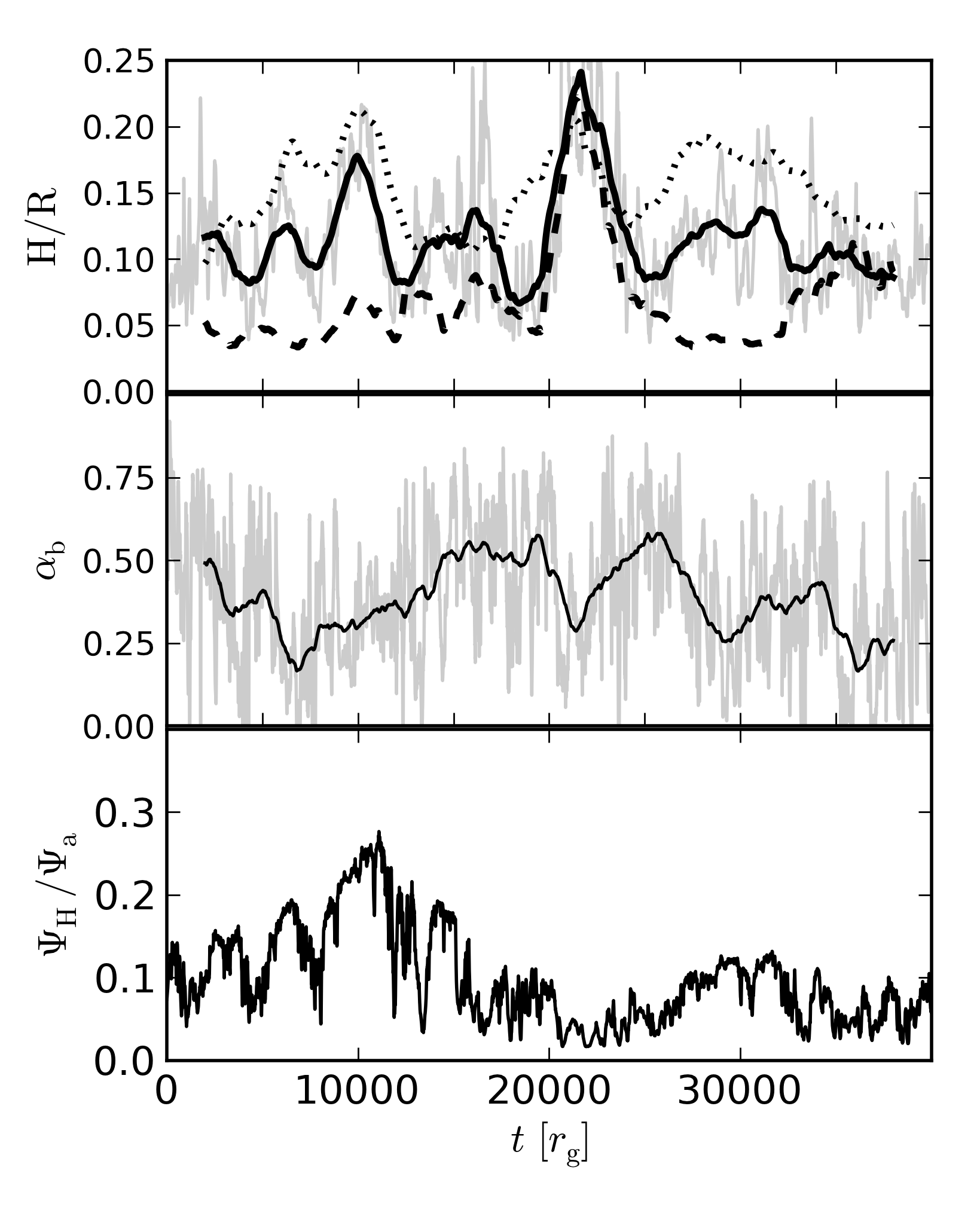}
\caption{Same as Figure \ref{fig:quantsVsT} but with values for our simulation RADLR.} 
\label{fig:quantsVsTLR}
\end{figure}

\begin{figure}
\centering
\includegraphics[width=3.2in,clip]{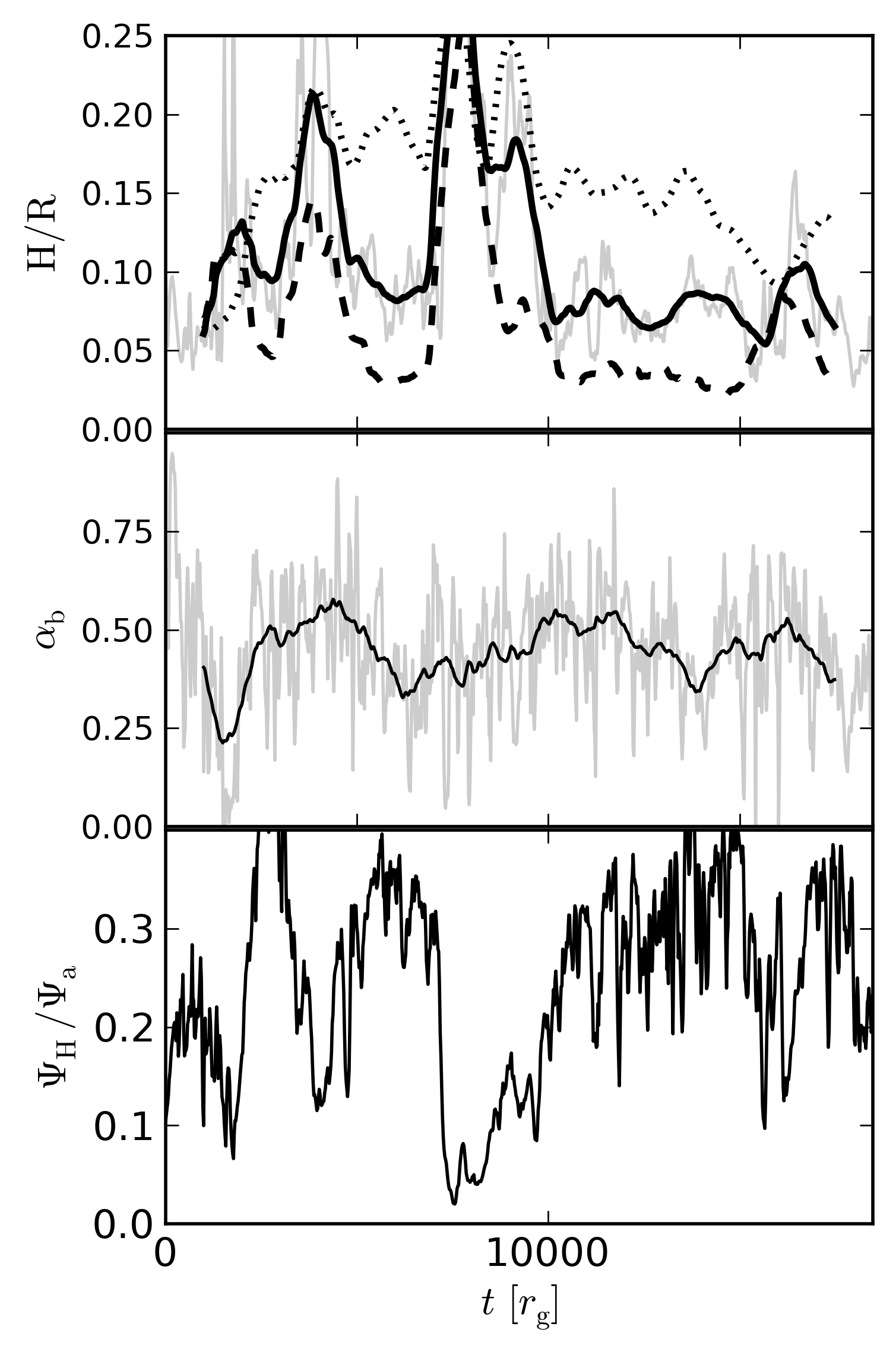}
\caption{Same as Figure \ref{fig:quantsVsT} but with values for our simulation RADvHR.} 
\label{fig:quantsVsTvHR}
\end{figure}

The time evolution of the disk scale-height, stress and magnetic flux presented in Figure \ref{fig:quantsVsTLR} for RADLR and Figure \ref{fig:quantsVsTvHR} for RADvHR at three different locations shows that, as expected, RADLR has more difficulty reaching a quasi-steady state value where a peak close to $t\approx$22,000$R_g/c$ in the disk scale-height has been observed and the values seem to be try to reach a quasi-steady state value at the end of the simulation. RADvHR presents strong variability and large peaks in the disk scale-height in the initial evolution and after 10,000$R_g/c$ the values at the three locations seem to be trying to converge to a constant value. 

In all of our three simulations, RADHR, RADLR, and RADvHR, the time evolution of the disk scale-height presented in Figures \ref{fig:quantsVsT}, \ref{fig:quantsVsTLR} and \ref{fig:quantsVsTvHR} at three different radii indicate that the disks are kept stable against the thermal collapse from radiative loses, and in all cases the value of $H/R$ seems to be converging to values around 0.1. A further study is required in order to fully determine whether the disk is stable or not. 

\begin{figure}
\centering
\includegraphics[width=3.2in,clip]{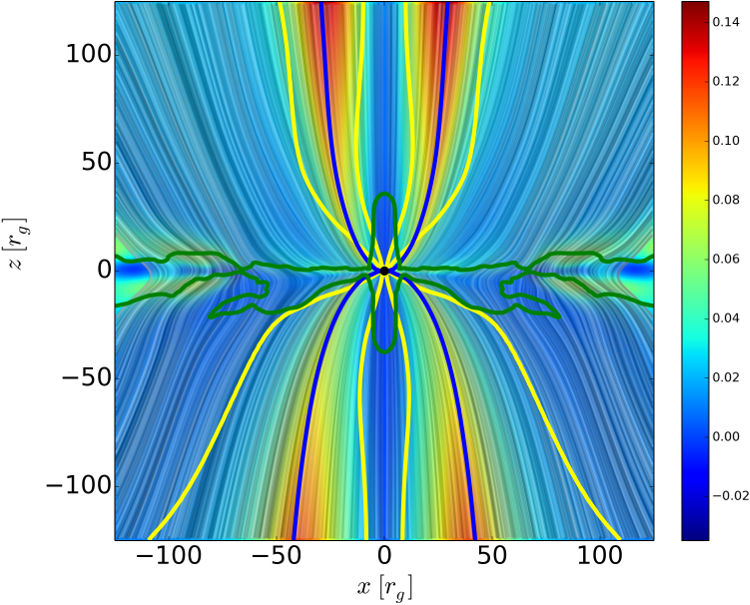} 
\includegraphics[width=3.2in,clip]{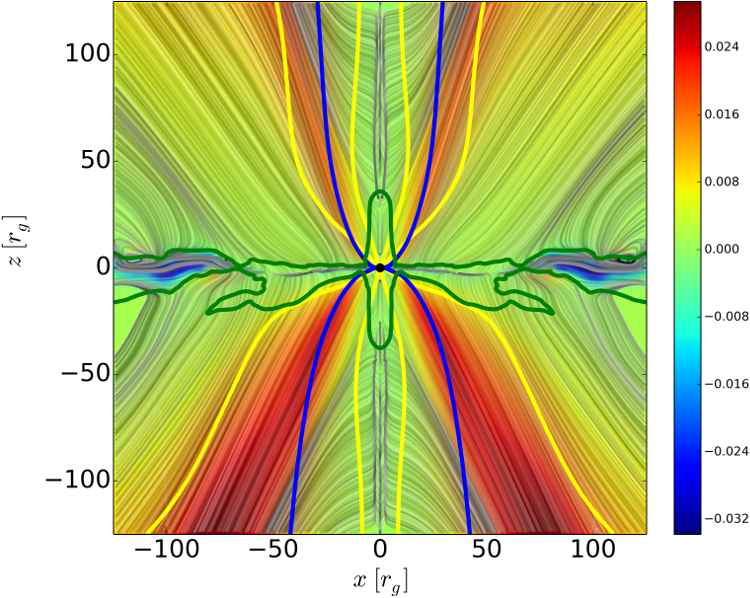}
\caption{Same as Figure \ref{fig:quantsVsT} for our simulation RADvHR.}
\label{fig:fig2a-vHR}
\end{figure}

The electromagnetic and radiation fluxes presented in Figure \ref{fig:fig2a-vHR} show that the EM luminosity has a more uniform angular, and more north-south symmetric, distribution, for RADvHR, with the jet emission in the north stronger when compared with the south direction and our simulation RADHR. Different from RADHR the radiation luminosity of RADvHR is broader and stronger in the south direction, the wind in RADvHR presenting a stronger emission in this case (yellow streams) with the $\tau_{es}$=1 surface showing in this simulation a broader opening angle in both directions indicating that at this time period a significant amount of material is blocking the radiation and similarly to RADHR most of the emission is coming from the region inside 15$R_g$.

\section{Discussion}
\label{sec:discuss}

We have presented the first GR-Rad-MHD simulations of a thin accretion disk around a spinning black hole in the so called MAD state with the objective to study the accretion efficiencies and compare our results with the recent paper from \citet{Avara16}. Our results have shown that the wind efficiency is greater than the jet efficiency, and the total efficiency reported in this work is similar to that found by \citet{Avara16} when taking into account that in their simulation the total efficiency is slightly higher due to the use of an \textit{ad hoc} cooling function that does not take into account the effects of scattering and absorption of the radiation. As in \citet{Avara16}, we have also seen the presence of a weak jet with a non-uniform distribution in the south direction indicating that a large amount of material in this region is blocking the radiation as seen by the location of $\tau$=1 for electron scattering. Our results can also be compared with the findings of \citet{Sadowski16c} whom computed the accretion efficiencies in a similar configuration however with the magnetic field in the SANE state and found accretion efficiencies close to the expected values of the NT73 model with the extra efficiencies coming from the plunging region. 

We quantify the effects of resolution on our results by comparing our fiducial simulations to both lower and high resolution simulations to test convergence.  Our simulation RADvHR demonstrates the highest efficiency from all of our RAD simulations since the magnetic flux transport has been more efficient, though MADiHR had an even larger efficiency. All simulations reported in this study have shown that the wind carries the largest portion of outgoing energy, and that the accretion flow can extract rotational energy from the black hole producing a weak jet, both results consistent with the findings of \citet{Avara16}. 

We have also carried out initial analyses of the stability of our thin disk and in all simulations we have found that our thin disk in the MAD state has been supplied with a large amount of magnetic flux which produced a strong magnetic pressure that was able to support the disk against the collapse due to radiative loses. \citet{Sadowski16c} through simulation in the SANE state found that stability was only reached in a configuration for the magnetic that is quadrupolar while in the dipole case the disk has collapsed. Our findings shows that a dipole configuration in the MAD state provided magnetic pressure strong enough to kept the disk stable. 

In a future study we will evolve our simulation RADvHR much longer in order to study the jet power, if the MAD state is reached further out, check if the absorption in the south direction will increase and also include different opacities like that used by \citet{Jiang16,McKinney16} to study how these opacities could affect the jet power, the build up of the MAD state and whether or not they have some effect in the disk stability.

\section{Conclusions}
\label{sec:conclusions}

We have shown the results of the first known simulations of
radiatively efficient thin disks, with $H/R\sim$0.1, around a spinning
black hole with $a/M$=0.5, that are in the MAD state and include
self-consistent radiative transport. Our longer high-resolution
simulation RADHR reached the MAD state out to 16$R_g$ and most energy
extracted by the disk is carried away by the wind consistent with the
findings of \citet{Avara16} but with a smaller total efficiency due to
a significantly smaller radiation component. This is because our
calculations include self-consistent absorption and scattering of the
radiation and advection by plunging material. We
measured the electromagnetic and radiation luminosities and we found
anisotropic emission in both cases. Radiation, on the other
hand, prefers emission to the north. A key result of this work that
will be explored in the future is the thermal stability of the disk,
likely a result of the very strong magnetic fields intrinsic to the
MAD state. It remains to be shown if there is a qualitative difference
in disk stabilization by a strong magnetic field in a standard disk
that is MRI dominated, compared to a disk in the MAD state where the
MRI no longer dominates the accretion process.

\section{Acknowledgments}

DMT thanks Ramesh Narayan, Chris Fragile, Olek Sadowski, Roman Gold, Jane Dai and Peter Polko for the several helpful comments. DMT thanks the Brazilian agencies CNPQ (proc 200022/2015-6), CAPES (88887.130860/2016-00) and FAPESP (2013/26258-4) for the financial support that made this project possible. DMT also thanks the Physics and Astronomy departments of University of Maryland  for their hospitality where the major part of this work has been carried out. MJA acknowledges partial support from NSF grants AST1028087, AST-1516150, PHY-1305730, PHY-1707946, OAC-1550436 and OAC-1516125, in addition to NASA/NSF/TCAN (NNX14AB46G), and use of the Deepthought2 cluster at UMD. JCM thanks the support by NASA/NSF/TCAN (NNX14AB46G) as well as computing time from NSF/XSEDE/TACC (TG-PHY120005) and NASA/Pleiades (SMD-14-5451).

\label{lastpage}
\end{document}